\documentclass[11pt]{article}
\usepackage{jheppub}
\usepackage{amsmath,amssymb,amsfonts,graphicx,slashed,amsthm,mathtools, enumerate, tensor, subfig}
\usepackage[dvipsnames]{xcolor}
\usepackage{arydshln}

\usepackage{comment}
\usepackage{hyperref}
\usepackage[utf8]{inputenc}
\usepackage[titletoc]{appendix}

\usepackage{braket}
\usepackage{color}
\usepackage{transparent}
\usepackage{MnSymbol}

\usepackage{tikz}
\usetikzlibrary{decorations,decorations.pathmorphing}
\usetikzlibrary{calc}
\usetikzlibrary{patterns}

\newcommand{\ft}{\text{flat}}

\newcommand{\AM}[1]{#1}
\newcommand{\AMM}[1]{#1}

\newcommand{\ha}{\hat{a}}
\newcommand{\bk}{\mathbf{k}}
\newcommand{\tin}{\text{in}}
\newcommand{\tout}{\text{out}}
\newcommand{\dt}[1]{\delta^{(3)}(#1)}
\newcommand{\tk}{\tilde{k}}
\newcommand{\dk}
{d\tilde{k}}
\newcommand{\al}[1]{z(#1)}

\newcommand{\md}[2]{\mathcal{D}{#1}{\mathcal{D}}{#2}}
\newcommand{\alf}{{\alpha}_f}
\newcommand{\ali}{\bar{\alpha}_i}
\newcommand{\celint}[2]{\int\!\frac{d{#1}}{2\pi i}d^2{#2}}
\newcommand{\ra}[1]{region $\mathcal{A}^{#1}$}
\newcommand{\rd}{region $\mathcal{D}$}
\newcommand{\dd}{\text{dd}}

\title{Mixed boundary conditions and Double-trace like deformations in Celestial holography and Wedge-like holography}

\author[a]{Machiko Fukada}
\author[b,c]{\!, Akihiro Miyata}

\affiliation[\,a]{Graduate School of Arts and Sciences, the University of Tokyo, Komaba, \\ Meguro-ku, Tokyo 153-8902, Japan}
\affiliation[\,b]{Kavli Institute for Theoretical Sciences, University of Chinese Academy of Sciences, Beijing 100190, China}
\affiliation[\,c]{Center for Gravitational Physics and Quantum Information,
Yukawa Institute for Theoretical Physics, Kyoto University,
Kitashirakawa Oiwakecho, Sakyo-ku,
Kyoto 606-8502, JAPAN}

\emailAdd{fukada.phys@gmail.com}
\emailAdd{a.miyata@ucas.ac.cn}

\preprint{UT-Komaba/23-3}

\abstract{
According to the AdS/CFT dictionary, adding a relevant double-trace deformation $f\int O^2$ to a holographic CFT action is dual to 
imposing mixed Neumann/Dirichlet boundary conditions for the field dual to $O$ in AdS.
We observed similar behaviour in codimension-two flat space holographies. We consider deformations of boundary conditions in flat spacetimes under flat space codimension-two holographies, celestial holography and Wedge-like holography.
 In the former celestial-holographic approach, we imposed boundary conditions on initial and final bulk states in the scattering. 
 We find that these non-trivial boundary conditions in the bulk induce 
   ``double deformations" on the Celestial CFT side, which can be understood as an analogy of double trace deformations in the usual AdS/CFT. We compute two-point bulk scattering amplitudes under the non-trivial deformed boundary conditions. 
 In the latter Wedge-like holography approach, we consider mixed Neumann/Dirichlet boundary conditions on the null infinity of the light-cone. We find that this mixing induces a renormalization flow in the dual Wedge CFT side under the Wedge holography, as in the usual AdS/CFT.
We argue that the discrepancy between the Wedge two-point function and the Celestial two-point function originates from a sensitivity of bulk massless fields to a regularization parameter to use the usual AdS/CFT techniques.
 } 


\begin{document}

\maketitle

\parskip=10pt


		
	

\section{Introduction and main results}

\paragraph{Condimension-two holographies of flat spacetime}

The holographic principle has  been a proposal that $(d+1)$-dimensional quantum gravitational theory can be described by a $d$-dimensional non-gravitational theory \cite{tHooft:1993dmi,Susskind:1994vu}. The AdS/CFT correspondence \cite{Maldacena:1997re} is a concrete example of this principle, where a quantum gravity in \AM{$(d+1)$-dimensional anti-de Sitter (AdS)} spacetime is described by a \AM{$d$-dimensional conformal field theory (CFT)} living at the AdS boundary, i.e. codimension-one holographic correspondence. 

Recent years,  \emph{codimension-two flat space holographies} have appeared, in which asymptotic flat quantum gravity is described by a two-dimensional lower  non-gravitational theory.
 A known codimension-two flat space holography is Celestial Holography \cite{Pasterski:2016qvg}. Celestial holography proposes that $n$-point scattering in asymptotic flat space in the boost basis is equivalent to $n$-point function in the dual \AM{CFT \cite{deBoer:2003vf,Pasterski:2016qvg,Cheung:2016iub,Kapec:2014opa,Donnay:2018neh} 
 (Celestial CFT, CCFT).}  In particular, massless case has been intensively investigated (As a summary of the literature of studies of massless  Celestial holography, see \cite{Pasterski:2021raf}). This holography allows us to \AM{capture }\cite{Fotopoulos:2020bqj,Strominger:2017zoo,He:2015zea,Pasterski:2017ylz,Guevara:2019ypd,Dumitrescu:2015fej} bulk asymptotic symmetries by the dual CFT. However, there are still many unclear points regarding the concreteness of the correspondences of Celestial holography. For example, we did not have a clear understanding of how the dynamics of the asymptotically flat spacetime of the bulk are related by the two-dimensional Celestial CFT until \AM{recently }\cite{Pasterski:2022lsl,Costello:2022jpg}.

 
 


There is another flat space holography, an application of Wedge holography \cite{Ogawa:2022fhy}. In this holography, the usual AdS/CFT techniques are used by dividing the Minkowski spacetime into the outside region with dS patch and the inside region with Euclidean AdS patch of the lightcone\footnote{In \cite{deBoer:2003vf}, this sort of treatment of the flat spacetime was suggested.}. 


\paragraph{Purpose of this paper:}
In this paper, we investigate boundary conditions in the \AM{four-dimensional} flat spacetime and consider deformations of the boundary conditions in two approaches: 1. Conventional celestial holography and 2. Wedge-like approach to celestial holography. We will explain in detail below;

\paragraph{Approach 1: The (initial and final) boundary conditions in celestial holography} 
 We will show in section \ref{sec:celestial dd} that changing asymptotic boundary conditions in the \AM{four-dimensional} asymptotically flat spacetime are related to the change of the celestial CFT.
In this paper, we give examples of non-trivial but simple asymptotic conditions, which give us a non-trivial celestial CFT interpretation. Such a non-trivial operator that acts on asymptotic states has to non-trivially change a vacuum and also be beyond Fock states. 
One simple example is gauge dressing. We newly propose another simple and non-trivial deformation in bulk: double deformation which adds $\exp\int (a_{\tin}^\dagger)^2$ to the in-vacuum, and $\exp\int (a_{\tout})^2$ to the out-vacuum.
We will introduce coherent path integral as a convenient tool to capture the boundary conditions from asymptotic states.

\paragraph{Approach 2: Mixed-boundary conditions in the wedge-like approach} 


We will show in section \ref{sec:Wedge-like hologrpahy} that the mixed Neumann/Dirichlet boundary conditions in wedge-like holography non-trivially change bulk two-point functions.
 In the $\rm{AdS}_{d+1}/\rm{CFT}_{d}$, this mixed boundary condition of a bulk field on $\rm{AdS}_{d+1}$  induces the (relevant) double-trace deformation in $\rm{CFT}_{d}$ \cite{Witten:2001ua,Gubser:2002zh,Gubser:2002vv,Aharony:2001pa,Aharony:2001dp,Berkooz:2002ug,Mueck:2002gm,Minces:2002wp,Petkou:2002bb,Sever:2002fk,Barbon:2002xk,Aharony:2005sh,Hartman:2006dy}\footnote{In usual $\rm{AdS}_{d+1}/\rm{CFT}_{d}$, the relevant double-trace deformations of the form $f \mathcal{O}^{2}$, where $\mathcal{O}$ is a single trace operator with the conformal dimension $d/2-1\leq 
\Delta<d/2$ induces the RG flow from the original (unitary) CFT to another (unitary) one in IR.
Along the RG flow, the conformal dimension $\Delta$ in UV is changed to $d-\Delta$ in IR.
Importantly, both conformal dimensions $\Delta,d-\Delta$ satisfy the unitarity bound, and this is consistent with the fact that CFT dual to AdS is unitary. 
}. 
In this paper, we investigate, in the wedge-like holography, whether the mixed Neumann/Dirichlet boundary conditions have a similar interpretation of an RG flow in AdS/CFT.

\paragraph{Main results}
In this paper, as explained above, we study how mixed boundary conditions in the bulk change in the dual Celestial and Wedge CFT. We show the results below for each holography;

\begin{itemize}
	\item \underline{For approach 1:celestial holography in section \ref{sec:celestial dd}}, we give examples of non-trivial boundary conditions mixing creation and annihilation eigenvalues (usually, the vacuum state corresponds to vanishing annihilation eigenvalue) at future and past infinities, respectively. 


 The dual CCFT induced by the mixed boundary conditions has the similar structure as the double-trace deformations in AdS/CFT, though not the same one  since celestial holography does not have the large-$N$ structure unlike the AdS/CFT.  We confirmed that the double-deformed celestial CFT two-point functions match with the bulk deformed two-point functions.\par 
We further examined a deformation of CCFT of the form: normal ordered $\exp[f\int O_iO_f]$, similar to the \textit{non-local} double-trace deformation in the AdS/CFT. For a massless case, the original delta-functional behaviour was simply strengthened or weakened by the sign of the parameter $f$.
%
	\item \underline{For approach 2: Wedge-like holography in section \ref{sec:Wedge-like hologrpahy}}, we examined the deformed Wedge-like CFT which is dual to the mixed boundary conditions on the null infinity of the light-cone.
   We introduced an IR regularization parameter to conformal dimensions to make use of the usual AdS/CFT techniques. From the behaviour of the bulk massless scalar two-point function, we find that, the mixed boundary conditions trigger an RG flow on the dual CFT side, from renormalized IR fixed point which corresponds to the Wedge CFT, to UV fixed point which corresponds to the Celestial CFT, if we look just at a massless scalar field.
 

 

 
\end{itemize}

\paragraph{Organization of this paper}
The paper is organized as follows.
 First, in section \ref{sec:operatorcorresp.}, we briefly introduce the notions of codimension-two flat space holographies and explain celestial holography and our Wedge-like formulation.
 In the next section \ref{sec:celestial dd}, we study mixed boundary conditions in the celestial holography.
  In section \ref{sec:Wedge-like hologrpahy}, we explain the Wedge-like holography and study the mixed Neumann/Dirichlet boundary conditions. 
  In the last section \ref{sec:conclusion}, we give some discussions  and list future directions.
 In appendix \ref{appendix:coherent}, we explain a formulation of celestial holography using coherent path integrals as preparation for a systematic treatment of double deformations in Celestial Holography.
 In appendix \ref{appendix:cfunction}, we evaluate a c-function under the Wedge-holography.
 In appendix \ref{appendix:harmonic basis}, we explain a method for solving boundary conditions appearing in the body of this paper.
 In appendix \ref{appendix:holograEOW}, we give an example with certain exotic boundary conditions comparing the body of this paper. Under the exotic boundary conditions, a two-point function, which is evaluated by using the Wedge-like formulation with some modifications, reproduces an expected celestial CFT result. 
 In appendix \ref{appendix:massiveScalar}, we give 
 a brief discussion of a massive scalar field case under the Wedge-like holography.

\section{Preliminary: Codimension-two flat space holographies}\label{sec:operatorcorresp.}
We will compute double deformations which are similar to double-trace deformation in AdS/CFT  (we will define it in \ref{sub: dd in CCFT}\AM{)} under  Minkowski/CFT correspondences (codimension-two flat space holography) in the next section. Na\"ively, we cannot apply the AdS/CFT correspondence directly to the Minkowski space and so it seems impossible to have a CFT description from Minkowski spacetime, since Minkowski spacetime has completely different structure from AdS spacetime.  
 Yet a possibility of applying (A)dS/CFT correspondence to the Minkowski spacetime by foliating properly was first pointed out by de Boer and \AM{Solodukhin }\cite{deBoer:2003vf,Solodukhin:2004gs}. More than a decade later the idea has developed to give many arguments of correspondence between gauge theories in \AM{four-dimensional} Minkowski spacetime and \AM{two-dimensional} CFTs on Celestial sphere living on asymptotic infinity, by either using the (A)dS/CFT \AM{correspondence }(e.g.\cite{Cheung:2016iub}) which we call \textit{Wedge-like holography}  or by not explicitly using it which we call \textit{Celestial holography}.\par 
  Which flat space holography should we choose for our purpose of deforming the theory by double deformation? The double-trace deformation have been defined and used to non-abelian gauge theories in the context of AdS/CFT. Since we actually don't know much about Celestial CFT which contains different properties from usual CFT, it would be appropriate to adopt AdS/CFT-like holography. In particular, we focus on Wedge \AM{holography }\cite{Akal:2020wfl,Ogawa:2022fhy} in section \ref{sec:Wedge-like hologrpahy}, one application of AdS/CFT holography. The reason is that Wedge holography is defined at each regions, outside of the light-cone or inside of the light-cone, separately, and so \AM{it is} easier to analyse. Other than it, we study Celestial holography, a most standard flat space holography which is intensively investigated recent years.\par 
   For the above reasons, we reformulate flat space holographies in this section in an explicit form in preparation for introducing \AM{double deformations.} We first set up for our formulation, and then look \AM{at} how these flat space holographies are characterised. 
\subsection{Convention and Minkowski metric}
In this paper we consider \AM{four-dimensional} Minkowski space with metric signature $(-1,1,1,1)$\AMM{,
\begin{equation*}
	ds^{2}=-(dX^{0})^{2}+(dX^{1})^{2}+(dX^{2})^{2}+(dX^{3})^{2}.
\end{equation*}}
 It is splitted by the light-cone into three regions. Region $\mathcal{D}$ is outside of the light-cone and written by dS patch:
\begin{equation}
	X^\mu=(r\sinh t,~ r\cosh t\cos\theta,~r\cosh t\sin\theta\cos\phi,~r\cosh t\sin\theta\sin\phi), \label{eq:dScoor}
\end{equation}
and region $\mathcal{A}$ is inside of the light-cone and written by (Euclidean) AdS patch:
\begin{equation}
	X^\mu=(\eta\cosh\rho,~\eta\sinh\rho\cos\theta,~\eta\sin\rho\sin\theta\cos\phi,~\eta\sin\rho\sin\theta\sin\phi).
\end{equation}
The metric on \rd~ and on \ra{} ~is
\begin{align}
	ds^2&=-d\eta^2+\eta^2(d\rho^2+\sinh^2\rho~d\Omega^2),\\
	ds^2&=dr^2+r^2(-dt^2+\cosh^2 t~d\Omega^2),
\end{align}
respectively. The \AM{foliations} by fixed $\eta,~r$ have the \AM{three-dimensional} AdS and dS metrics. 
Region $\mathcal{A}$ is also splitted into region $\mathcal{A}^+:\eta>0$ and \ra{-}: $\eta<0$. In Wedge holography, the Wedge CFT is considered to live in a sphere of $\rho\to\infty$ or $r\to\infty$ limit on the EOW-branes $\eta=\eta_1,~\eta_2$ or $t=t_1,t=t_2$ which two-dimensional CFTs are dual to the bulk gravity regions between the two-branes within each regions: \ra{+}, \ra{-}, or \rd. 
\par 
On the other hand, Celestial operators in Celestial CFT live on the two-dimensional sphere called Celestial sphere embedded in particles' momenta. Parametrizing a massive and massless momentum as
\begin{align}
m\cdot{p^\mu}=&\frac{m}{2\omega}(1+{\omega^2}+z\bar{z},1-z\bar{z}-{\omega^2},z+\bar{z},i(\bar{z}-z))\\
 k^\mu=\omega\cdot q^\mu:=&\omega\frac{2}{1+\cos\theta_0}(1,\cos\theta_0,\sin\theta_0\cos\phi_0,\sin\theta_0\sin\phi_0)\\
 =&\omega(1 + z\bar{z}, 1 - z\bar{z}, \bar{z}+ z,  i(\bar{z}-z)),
\end{align}
the Celestial sphere in the momentum space and null infinity in the bulk space coincide. 
%
%
\subsection{Wedge-like holography's formulation}
Our starting point of the Wedge-like holography is that there exists an extrapolate dictionary between a bulk scalar in Minkowski space and a family of Wedge-like operators in Mellin basis. 
This starting assumption is motivated by a bulk reconstruction of Celestial operator proposed in \cite{Donnay:2018neh} and an extrapolate dictionary of Celestial operator subsequently proposed in \cite{Pasterski:2021dqe} (see \AM{also \cite{Donnay:2022sdg,Iacobacci:2022yjo,Sleight:2023ojm})}.


\AMM{We define an extrapolate dictionary in Wedge-like holography as follows; first we decompose the bulk field $\Psi$ into modes labeled by $\Delta$, 
\begin{equation*}
    \Psi=\int\frac{d\Delta}{2\pi i} \Psi_{\Delta} ,
\end{equation*}
then we define the operator $O_W^\Delta$ by 
\begin{equation*}
	\lim_{\rho\to \infty} e^{-(\Delta-2)\rho }  \Psi_{\Delta} = O_W^\Delta.
\end{equation*}}
We will see the explicit construction in section \ref{sec:Wedge-like hologrpahy}.


In the section \ref{sec:Wedge-like hologrpahy}, we will derive a GKPW-like dictionary
\AMM{\begin{equation}
Z_\ft[\Psi|_{\text{bdy}}=\Psi_b]/Z_\ft[\Psi|_{\Psi_b=0}]=\left\langle \exp\left[{e^{2\epsilon R}\int\! d^2w\frac{d\Delta}{2\pi i}~O^W_{\Delta^*}(w,\bar{w})\Psi^\Delta_b(w,\bar{w})}\right]\right\rangle_\text{CFT}/\langle 1\rangle_{\text{CFT}},\label{GKPW}
\end{equation}}
following the proof used in \cite{Terashima:2017gmc}\footnote{Since there is a subtle problem regarding the discrepancy between the GKPW and the extrapolate \AM{dictionary }\cite{Harlow:2011ke}, the proof of the GKPW is limited to Wedge holography within \ra{+}. } from the extrapolate dictionary, where we take the boundary conditions at $\rho=R\gg 1$.


We will also consider the holography beyond Wedge holography. We will call this holography Wedge-like holography. Here we will clarify the difference between them. In Wedge holography, one consider regions within wedges in each splitted regions. On the other hand, the Wedge-like holography is considered to across wedge boundaries.

\subsection{Celestial holography}\label{subsec:Celestialholography}
We briefly introduce a \AM{two-dimensional} operator living in the celestial sphere in a framework of celestial holography. Given a \AM{four-dimensional} bulk free scalar field $\Psi$, we can define  the celestial operator $\hat{O}_\Delta(w,\bar{w})$ on the principal continuous series $\Delta\in 1+i\mathbb{R}$  by \cite{Donnay:2020guq}
\begin{equation}\label{eq:innerProduct}
   \begin{split}
        \hat{O}_\Delta^\pm=&(\Psi(X^\mu),\phi^\pm_\Delta(X;w,\bar{w}))_\Sigma\\
        &=\int_\Sigma d^3X[\Psi(X^\mu)\partial_0{\phi_\Delta^\pm}^*(X;w,\bar{w})-{\phi^\pm_\Delta}^*(X;w,\bar{w})\partial_0\Psi(X^\mu)]
   \end{split}
\end{equation}
using a wavefunction called conformal primary wavefunction $\phi_\Delta$ we will introduce below with the Klein-Gordon inner product $(,)$. 
$\phi_\Delta$ are the \AM{conformal} primary wavefunctions defined by
\begin{equation}
		\phi^\pm_\Delta(X_\pm; w,\bar{w})=\int^\infty_0 d\omega~\omega^{\Delta-1}e^{\pm ikX_\pm}=\frac{\Gamma(\Delta)}{(\pm i)^\Delta}\frac{1}{(-q\cdot X_\pm)^\Delta}.\label{def:cpw}
\end{equation}
We introduce the regulator $X^\mu_\pm=X^\mu\mp i\kappa(1,0,0,0),~\kappa>0$ to change the path of the integration depending on the sign. 
We usually take this product to be the Klein-Gordon inner product on a Cauchy slice $\Sigma$. This operator does not depend on the choice of the Cauchy slice as long as these satisfy the Klein-Gordon equation. Conversely, we can reconstruct the free bulk scalar field using the celestial \AM{operator }\cite{Donnay:2018neh} as we will see below. We also see the statement of Celestial holography.


\paragraph{massless case}
A free massless scalar \AM{field} $\Psi$ can be written as 
\begin{equation}
	\begin{split}
		\Psi(X)&=\int\! d\tilde k ~\left[a(\bk)e^{ik\cdot {X}}+a^\dagger(\bk)e^{-ik \cdot {X}}\right]\\
 &=\int\! d^2 w\int_{1-i\infty}^{1+i\infty}\!\frac{d\Delta}{2\pi i}  \left[{\hat{O}}^+_{\Delta}(w,\bar{w})\phi^+_{\Delta^*} (X;w,\bar{w})+\hat{O}_{\Delta}^-(w,\bar{w})\phi^-_{\Delta^*} (X;w,\bar{w})\right]\qquad \label{eq:mode exp}
	\end{split}
\end{equation}
where $d\tk$ is the Lorentz invariant measure
\begin{align}
	d\tk:=\frac{d^3\bk}{2k^0}\label{lorents invarian measure}.
\end{align}
Celestial holography denotes that, for asymptotically free theory, Celestial CFT operators are related to asymptotic fields $\Psi_{\tin(\tout)}$ as
%
\AMM{\begin{align}
    \hat{O}^{\Delta}_i(w,\bar{w})&=\int d\omega~\omega^{\Delta-1}\ha^\dagger_\tin(\bk(\omega,w,\bar{w}))\label{operator corresp. o_f a_f}\\
    \hat{O}^\Delta_f(w,\bar{w})&=\int d\omega~\omega^{\Delta-1}\ha_\tout(\bk(\omega,w,\bar{w})).\label{operator corresp. o_i a_i}
\end{align}}
\AMM{Here, we note that although the creation and annihilation operators are originally defined at the space-like Cauchy slice at $X^{0}\to \pm \infty$, i.e., $\Sigma_{X^{0}\to \pm \infty}$, we can push the space-like Cauchy slices to null infinities $\mathcal{I}^{\pm}$ for massless fields as shown in appendix A of \cite{Donnay:2020guq}. Therefore, we can consider their operators as being defined at null infinities;
\begin{equation}
	\begin{aligned}
		\ha^\dagger_\tin(\bk) & = \left(\Psi(X^\mu),e^{ik\cdot {X}} \right)_{\Sigma_{X^{0}\to -\infty }} = \left(\Psi(X^\mu),e^{ik\cdot {X}} \right)_{ \mathcal{I}^{-} } \\
		\ha_\tout(\bk) & = \left(\Psi(X^\mu),e^{-ik\cdot {X}} \right)_{\Sigma_{X^{0}\to \infty }} = \left(\Psi(X^\mu),e^{-ik\cdot {X}} \right)_{ \mathcal{I}^{+} }.
	\end{aligned}
\end{equation}
Correspondingly, the celestial CFT operators $\hat{O}^{\Delta}_i(w,\bar{w}),\hat{O}^\Delta_f(w,\bar{w})$ for the massless field can be regarded as operators defined at null infinities $\mathcal{I}^{\pm}$.}

\AMM{These operators} \AM{reproduce} the statement
\begin{align}
    \langle \hat{O}_f^{\Delta_1}(w_1,\bar{w}_1),\dots \hat{O}^{\Delta_n}_i(w_n,\bar{w}_n)\rangle=\int d\omega_1\dots d\omega_n~\omega^{\Delta_1-1}_1\dots\omega^{\Delta_n-1}_n\mathcal{A}_n(\bk_1,\dots,\bk_n).
\end{align}
Especially, a two-point function in free theory is
	\begin{equation}
		\langle \hat{O}_f^{\Delta}(z,\bar{z})\hat{O}_i^{\Delta'}(z',\bar{z'})\rangle=2\pi i\delta(\Delta+\Delta'-2)\delta^{(2)}(z-z')=\int^\infty_0 d\omega d\omega'~\omega^{\Delta-1}\omega'^{(\Delta'-1)}2k^0\delta^{(3)}(\omega \mathbf{q}+\omega'\mathbf{q}').\label{eq:CCFTtwopointfunction}
	\end{equation}
%
%
\paragraph{massive case}
A massive asymptotic scalar free field $\Psi$ with \AMM{time $X^{0}\to\pm\infty$}
 in flat space is decomposed by the principal continuous series $\Delta\in 1+i\mathbb{R}_>$ to extract the Celestial CFT operator $\hat{O}_\Delta$ \cite{Donnay:2018neh}
\begin{align}
	\Psi(X)&=\int \!d^2w \int^{1+i\infty}_1\!\frac{d\Delta}{2\pi i} \left[N_\Delta \hat{O}_{\Delta}(w,\bar{w})\phi^+_{\Delta^*} (X;w,\bar{w})+N_\Delta \hat{O}^\dagger_{\Delta}(w,\bar{w})\phi^-_{\Delta^*} (X;w,\bar{w})\right],\\
 \quad N_\Delta&=m^2\frac{|\Delta-1|^2}{4\pi^3(2\pi)^3},
\end{align}
where the conformal primary wavefunctions are defined by
\AM{\begin{align}
	\phi_\Delta(X_\pm; w,\bar{w})&=\int^\infty_0 \frac{d^3p^i}{p^0}~G_\Delta(p(\omega,z,\bar{z});w,\bar{w})e^{\pm impX_\pm} \notag\\
	&=\int \frac{d\omega}{\omega^3}\int d^2z~ G_\Delta(\omega,z,\bar{z};w,\bar{w})e^{\pm imp(\omega,z,\bar{z})X_\pm} \label{eq:cpwForMassive} \\
	&\propto
	\begin{cases}
		\frac{4\pi}{im}\frac{\pi}{2}(\mp i)^{\Delta}\frac{\sqrt{-X^2}^{\Delta-1}}{(-X_\pm\cdot q)^\Delta}H^{(\pm)}_{\Delta-1}(m\sqrt{-X^2})&\text{AdS region},\\
		\frac{4\pi}{im}\frac{\sqrt{-X^2}^{\Delta-1}}{(-X_\pm\cdot q)^\Delta}K_{\Delta-1}(m\sqrt{X^2}) & \text{dS region}.
	\end{cases} \notag
	\end{align}}
In massive cases, celestial holography becomes
\begin{equation}
	\langle \hat{O}_{\Delta_1}(w_1,\bar{w}_1)\dots \rangle=\prod_{i=1,\dots,n}\int^\infty_0 \frac{d\omega_i}{\omega_i^3}dz_id\bar{z}_i\frac{1}{(-p(\omega,z,\bar{z})\cdot q(w,\bar{w}))^{\Delta_i}}A(p_i(\omega,z,\bar{z})).
\end{equation}
Especially, two-point functions \AM{are \cite{Costa:2014kfa}
\begin{align}
	\langle \hat{O}^{out}_{\Delta}(w,\bar{w})\hat{O}^{in}_{\Delta'}(w',\bar{w'})\rangle
	=&\int^\infty_0\frac{d\omega}{\omega^6} dz d\bar{z}\frac{2\omega^3}{m^3}\left(\frac{\omega}{\omega^2+|z-w|^2}\right)^\Delta\left(\frac{\omega}{\omega^2+|z-w'|^2}\right)^\Delta \\
	=&\frac{2\pi}{m^2(1-\Delta)}\delta(\Delta-\Delta')\frac{1}{|w-w'|^{2-\Delta}}-\frac{2\pi^2}{m^2(\Delta-1)^2}\delta(\Delta+\Delta'-2)\delta^{(2)}(w-w'). \label{eq:massiveCCFTtwopoint}
\end{align}}
Because we restrict $\Im[\Delta]$ to non-negative values, only the first term remains. \\\\
\underline{A comment about ordering in Celestial CFT}:
 In usual CFT, normal ordering is determined by radius, a continuous parameter.  \AMM{In Celestial CFT, the situation is completely different. There are only two-types of operators, in- and out- operators, in accordance with in-coming particles or out-going particles in the dual bulk.  In the spirit of Celestial holgraphy that a scattering amplitude corresponds to a correlation function of celestial operators, the ordering in Celestial CFT has to be related with the bulk time ordering. Through the Mellin transformation of bulk scattering amplitudes, we have just two types: in- and out- operators $\hat{O}_i(w,\bar{w}),\hat{O}_f(w,\bar{w})$, since a scattering is completely determined by asymptotic states at past- and future- infinities. In-operators $\hat{O}_i(w,\bar{w})$ are always to be ordered \AM{to} the left of out-operators $\hat{O}_f(w,\bar{w})$}.
%
%
\section{Double deformation for Celestial holography}\label{sec:celestial dd}
In this section, we will see that we can observe a local double-trace-like behaviour in Celestial CFT which we call ``\textit{double deformation}" in this paper, as the AdS/CFT holography, if we impose mixed boundary condition on asymptotic fields at \AMM{$X^{0}\to\pm\infty$}. 
In subsection \ref{subsec:bdy cond in CCFT}, we formulate the mixed asymptotic conditions using coherent representation and show the dual Celestial CFT description is local double-trace-like deformation. As a byproduct by the coherent path integration, we obtain a double-traced Celestial CFT's Lagrangian for scalar fields. In subsection \ref{subsec:two-point in CCFT}, we compute the double deformed two-point functions explicitly. An analogy of RG flow triggered by the double-trace deformation in AdS/CFT is unclear from them in this section. The interpretation as a role of RG flow will become clear in section \ref{sec:Wedge-like hologrpahy}. We also investigate on non-local double-trace like deformation in subsection \ref{subsec:non-local}. Non-local cases have many more applications in AdS/CFT. We finally examine \AM{local-double deformation} of conformal soft photon. We can thereby compare our results with known effective actions in Celestial CFT. 
 \par
%
%
%
For simplicity, \AM{we} focus on a scalar field. Although we mainly focus on the massless case here, we can discuss the massive case in the same way.

 \AMM{For notational convenience, in this section, we use $t$ as the time coordinate instead of $X^{0}$.}

\subsection{Asymptotic boundary conditions and local double  deformation}\label{subsec:bdy cond in CCFT}
In this subsection, we will show that changing initial and final asymptotic conditions leads to local double-trace-like deformation for Celestial CFT side. Here, it should be mentioned that coherent \AM{quantization }\cite{Zhang:1999is} is a useful tool for our purpose. Since $n$-point functions in Celestial CFT \AM{are} equivalent to scattering, coherent basis in the bulk allows us to easily move from the bulk to the Celestial path integral formulation. More importantly, we show that the mixed asymptotic conditions on asymptotic fields \textit{in coherent basis} 
\begin{equation}
    z_\tin=f_i\bar{z}_\tin,\quad \bar{z}_\tout=f_f z_\tout,\quad f\in \mathbb{R}
\end{equation}
\AM{lead} to double deformations, where $z$ is the eigenvalue of \AM{an annihilation operator} $\hat{a}$ by its ket state and $\bar{z}$ is the eigenvalue of \AM{a creation operator} ${a}^\dagger$ by its bra state as
\AMM{\begin{align}
	\ha (\bk,t)|z\rangle=z(\bk,t)|z\rangle,\quad \langle z |\ha^\dagger(\bk,t)=\langle z|\bar{z}(\bk,t).
\end{align}}
We can derive bulk scatterings under these boundary conditions by using the coherent path integral. For detailed coherent quantization and path integral formulation, see Appendix \ref{appendix:coherent}.\par 
In \ref{subsub:asymptotic condition}, we first investigate on mixed asymptotic conditions in coherent basis. Then in \ref{sub: dd in CCFT}, using Celestial holography, we deduce that changing asymptotic conditions amounts to inserting double-trace-like deformation on the side of \AM{Celestial} CFT. 
\par 
We remark that in the path integral, we have to take time-ordering $T$ which indicates that all $\hat{O}^\dagger_f$ operators should be always to the left of all $\hat{O}_i$ operators. On the other hand, in the CCFT, we take normal ordering $:~:$, which has the opposite ordering to $T$.
\subsubsection{Asymptotic conditions and asymptotic states}\label{subsub:asymptotic condition}
We show that \AM{the conditions
\begin{equation}
    z_\tin=f_i\bar{z}_\tin,\quad \bar{z}_\tout=f_f z_\tout,\quad f\in \mathbb{R}
\end{equation}}
correspond to the mixed asymptotic conditions
the double-deformed asymptotic states
\begin{align}
    |0\rangle_\dd=\exp\left\{\frac{f}{2}\int d\tilde{k}~(a^\dagger_\tin(\bk))^2\right\}|0\rangle. \label{def:double-trace deformation}
\end{align}
We call the correspondence between asymptotic condition in the state form and in the field form, asymptotic state-field correspondence. 

\paragraph{double deformation and changing asymptotic condition}
Suppose that instead of taking Fock states as asymptotic states, we take ``doubled multi-particle states" as asymptotic states:
\begin{align}
    |0\rangle_\dd=\exp\left\{\frac{f}{2}\int d\tilde{k}~(a^\dagger_\tin(\bk))^2\right\}|0\rangle. 
\end{align}
We will show that the mixed boundary conditions 
\begin{align}
    z_\tin(\bk)=f_i\bar{z}_\tin(\bk),\quad \bar{z}_\tout(\bk)=f_fz_\tout(\bk)
\end{align}
imply the above doubled multi-particle states.\par 
Letting the annihilation operator act on the doubled state as
\begin{equation}
    \begin{split}
        a_\tin(\bk)|i\rangle_\dd&=a_\tin(\bk)\exp\left\{\frac{f}{2}\int d\tilde{k}~(a^\dagger_\tin(\bk))^2\right\}|i\rangle\\
        &=fa_\tin^\dagger(\bk)|i\rangle_\dd
    \end{split}
\end{equation}
and inserting the complete set in coherent basis
\begin{align}
	\hat{I}=\int \frac{\mathcal{D}z \mathcal{D}\bar{z}}{ V}|z\rangle\langle z|, \quad 
	V=\prod_{\bk}(2\pi i)2\omega_\bk,
\end{align}
we have
\begin{equation}
\begin{split}
       \int \frac{\md{z}{\bar{z}}}{2\pi iV}f\bar{z}_\tin(\bk)|{z}_\tin\rangle\langle \bar{z}_\tin|i\rangle_\dd&=\int \frac{\md{z}{\bar{z}}}{2\pi iV}|z_\tin\rangle\langle \bar{z}_\tin|fa^\dagger_\tin(\bk)|i\rangle_\dd\\
       &=fa^\dagger_\tin(\bk)|i\rangle_\dd=a_\tin(\bk)|i\rangle_\dd\\
       &= \int \frac{\md{z}{\bar{z}}}{2\pi iV}a_\tin(\bk)|z_\tin\rangle\langle z_\tin|i\rangle_\dd\\
       &=\int \frac{\md{z}{\bar{z}}}{2\pi iV}{z}_\tin(\bk)|z_\tin\rangle\langle z_\tin|i\rangle_\dd.
\end{split}
\end{equation}
This condition is achieved if the following relation holds\footnote{Gauge \AM{dressing }\cite{Kulish:1970ut,Chung:1965zza,Hirai:2019gio,Choi:2019rlz,Hirai:2020kzx,Hirai:2022yqw,Choi:2017ylo,Kapec:2017tkm} also gives another boundary conditions: $z_\tin=const.,\quad \bar{z}_\tout=const.$.}
\begin{align}
    z_\tin(\bk)=f\bar{z}_\tin(\bk).
\end{align}

\paragraph{Schwinger's variational principle} 
We can deduce the above result in another way.
\AM{To this end, we use the stationary phase approximation \cite{Schwinger:1951ex,Marolf:2004fy} and the coherent path integral.}
First, we consider the case when $|i\rangle=|0\rangle$. Since the wavefunction in coherent basis in equation \eqref{path integral in coherent basis} is
\begin{align}
    \langle z|0 \rangle =e^{-\frac{1}{2}\int \dk~|z_\tin(\bk)|^2},
\end{align}
the stationary phase condition becomes
\begin{align}
    \left.\delta \left[e^{iS}\psi_i\psi^*_f\right]\right|_{\text{E.O.M.}}=&0\\
   i.e. \quad\delta\left|_{\text{E.O.M.}}\int\!\dk dt\right.~&\left[-\frac{1}{2}\left(\bar{z}(\bk,t)\dot{z}(\bk,t)-z(\bk,t)\dot{\bar{z}}(\bk,t)\right)\right]\\
  & -\frac{1}{2}|z_\tin(\bk)|^2-\frac{1}{2}|z_\tout(\bk)|^2=0\notag.
\end{align}
Thus, the asymptotic boundary condition becomes
\begin{align}
\bar{z}_\tout\delta z_\tout&=0,&z_\tin\delta \bar{z}_\tin&=0\\
    \to\quad \bar{z}_\tout(\bk)&=0,&z_\tin(\bk)&=0.\label{eq:bdycondVacum}
\end{align}
Let's now consider the dual state of double-trace deformed state \eqref{def:double-trace deformation}. Since
\begin{align}
    \psi_i&=\left\langle  z\left|\exp\left\{\frac{f_i}{2}\int d\tilde{k}~(a^\dagger_\tin(\bk))^2\right\}\right|0\right\rangle=\exp \left[\left\{-\int \dk~\frac{1}{2}|z_\tin(\bk)|^2+\frac{f_i}{2}(\bar{z}_\tin(\bk))^2\right\}\right],\\
     \psi^*_f&=\left\langle 0\left|\exp\left\{\frac{f_f}{2}\int d\tilde{k}~(a_\tout(\bk))^2\right\}\right|z\right\rangle=\exp \left[\left\{-\int \dk~\frac{1}{2}|z_\tout(\bk)|^2+\frac{f_f}{2}({z}_\tout(\bk))^2\right\}\right],
\end{align}
the asymptotic boundary condition becomes
\begin{align}
    (\bar{z}_\tout-f_fz_\tout z)\delta z_\tout&=0,&(z_\tin-f_i \bar{z}_\tin)\delta \bar{z}_\tin &=0\\
    \to \bar{z}_\tout-f_fz_\tout &=0,& z_\tin-f_i \bar{z}_\tin&=0.
\end{align}
\subsubsection{Double-trace like deformation in Celestial CFT}\label{sub: dd in CCFT}
As mentioned before, we can also observe a double-trace like behaviour in Celestial CFT as AdS/CFT. In the AdS/CFT, double-trace deformation means adding double-trace operator consisting of products of two single trace operators to the generating function.
We define double-trace like deformation in \AM{Celestial} CFT to add just products of two CCFT primary operators to \AM{the} generating function although the Celestial operators do not have large-$N$ structure unlike the AdS/CFT correspondence. 
 We will call this double-trace-like deformation  ``\emph{double deformation}".\par 
In the operator formalism,  using the bulk-reconstruction of Celestial CFT\AM{ \eqref{operator corresp. o_f a_f}, \eqref{operator corresp. o_i a_i}}
and $k^\mu=\omega(1+w\bar{w},\dots)$, the dual states in Celestial CFT of the bulk asymptotic states are
\begin{equation}
   \begin{split}
       \int \dk ~(a^\dagger_{\tin}(\bk))^2|0\rangle&=\int \dk~\left(\int\frac{d\Delta}{2\pi i}~\omega^{-\Delta}\hat{O}^\dagger_i\right)^2|0\rangle\\
       &=\int d^2w d\omega~\int\frac{d\Delta}{2\pi i}\frac{d\Delta'}{2\pi i}~\hat{O}^{\Delta}_i\hat{O}^{\Delta'}_i\omega^{1-\Delta-\Delta'}|0\rangle\\
       &=\int d^2w\frac{d\Delta}{2\pi i}\hat{O}_i^\Delta(w,\bar{w})\hat{O}_i^{\Delta^*}(w,\bar{w})|0\rangle.
   \end{split} 
\end{equation}
Thus, we find that double deformation in the bulk, and therefore mixing asymptotic conditions amounts to double-trace deformation in Celestial CFT. We can check this in the coherent path integral \AM{formulation
\begin{equation}
    \begin{split}
        Z^{f_i,f_f}=&\int\!\md{z}{\bar{z}}~\exp \left[-S_{\text{bulk}}+\frac{f_i}{2}\int \dk~{z_i^2(\bk)}+\frac{f_f}{2}\int \dk ~z_f^2(\bk)\right]\\
        =&\int\!\md{\bar{\alpha}}{\alpha}~\exp \left[-S_{\text{CCFT}}+\frac{f_i}{2}\int\frac{d\Delta}{2\pi i}d^2w~{|\bar{\alpha}^\Delta(w,\bar{w})|^2}+\frac{f_f}{2}\int \frac{d\Delta}{2\pi i}d^2w ~|\alpha^\Delta(w,\bar{w})|^2\right].
    \end{split}
\end{equation} }\par 
\subsection{two-point function for local double-trace-like deformation}\label{subsec:two-point in CCFT}
In this subsection, we compute the two-point functions from the bulk through the bulk path integral and also the two-point functions in celestial CFT. We confirm that these coincide.

\subsubsection{two-point function from the bulk path integral}\label{subsec:two-point function from the dual bulk}
We calculate the double deformed two-point function from the bulk path integral. The partition function is now written as
\begin{equation}
    \begin{split}
        Z^{\text{double-traced}}[J_f,J_i]&=\int\md{^2{\alpha}_i}{^2\alpha_f}~\langle \bar{\alpha}_i|i\rangle\langle f|\alpha_f\rangle \int \md{\bar{\alpha}(0)}{{\alpha}(0)}\left\langle \bar{\alpha}_f|\alpha(t_f)\rangle\langle\bar{\alpha}(t_i)|{\alpha}_i\right\rangle e^0\\
    &\propto\int\md{^2{\alpha}_i}{^2\alpha_f}~e^{\int J_f\alf+\bar{J}_i\bar{\alpha}_i+{\alpha}_i\bar\alpha_f-|\alpha_f|^2-|\alpha_i|^2+\frac{f_f}{2}\alpha_f^2+\frac{f_i}{2}\bar\alpha_i^2}.
    \end{split}
\end{equation}
Recalling that the deformed boundary condition is
\begin{align}
	\alpha_i=f_i\bar{\alpha}_i+\bar{J}_i,\quad \bar{\alpha}_f=f_f\alpha_f+J_f
\end{align}
and integrating by the two fields 
$\bar\alpha_i,~\alpha_f$ seen as real fields\AM{\footnote{We can of course obtain the same result by directly integrating by the two \textit{complex} fields.}}, we find that the Mellin transform of the deformed two-point scattering is
\begin{align}
	\langle \hat{O}_f^\Delta \hat{O}_i^{\Delta_i}\rangle=2\pi i \frac{1}{1-f_if_f}\delta(\Delta_i-\Delta_f^*)\delta^2(w-w'). \label{eq:two-pointMasslessS}
\end{align}
As a byproduct, we can obtain the Lagrangian of double-trace deformed Celestial CFT. Using the partition function becomes 
\begin{equation}
    \begin{split}
        Z^{\text{double-traced}}&
    \propto\int\md{^2{\alpha}_i}{^2\alpha_f}~e^{\int {\alpha}_i\bar\alpha_f-|\alpha_f|^2-|\alpha_i|^2+\frac{f_f}{2}\alpha_f^2+\frac{f_i}{2}\bar\alpha_i^2}\\
    &=\int\mathcal{D}\bar{\alpha}_i\mathcal{D}\alpha_f~e^{\int -\bar{\alpha}_i\alpha_f+\frac{f_f}{2}\alpha_f^2+\frac{f_i}{2}\bar\alpha_i^2},
    \end{split}
\end{equation}
we obtain the Lagrangian of double deformed Celestial CFT
\begin{align}
   { iL_{\text{CCFT}}}[\bar\alpha_i,\alpha_f]&=\int d^2w\frac{d\Delta}{2\pi i}\left[-\bar{\alpha}^{\Delta}_i\alpha^{\Delta^*}_f+\frac{f_f}{2}\bar\alpha_i^\Delta\bar\alpha_i^{\Delta^*} +\frac{f_i}{2}\alpha_f^{\Delta}\alpha_f^{\Delta^*}  \right].
\end{align}

While this Lagrangian has the information of the two point function \eqref{eq:two-pointMasslessS}, it looks trivial in the sense that the Lagrangian does not have derivatives of the field. 
This is a consequence of the absence of a large gauge symmetry for the bulk scalar field.
In contrast, if there is a large gauge symmetry, we can prepare asymptotic fields with bulk derivatives on Celestial sphere by choosing a suitable gauge in general.
\AMM{This would enable us to have a CFT Lagrangian including derivatives of asymptotic fields.}

\subsubsection{two-point function from the Celestial CFT }
We can deduce the same result from the celestial CFT formalism, the double deformed two-point function parametrized by $f_i$ is
\begin{equation}
	\begin{split}
		G^{f_f,f_i}&=\left\langle \exp\left\{\frac{f_f}{2}\int\!  \frac{d\Delta}{2\pi i}d^2w~\left|\hat{O}^{\Delta}_f\right|^2\right\}\hat{O}^{\Delta_f}_f (w_f,\bar{w}_f)\hat{O}^{\Delta_i}_i(w_i,\bar{w}_i)\exp\left\{\frac{f_i}{2}\int\!  \frac{d\Delta}{2\pi i}d^2w~|\hat{O}_i^\Delta|^2\right\}\right\rangle\\
		&\times\left\langle \exp\left\{\frac{f_f}{2}\int\!  \frac{d\Delta}{2\pi i}d^2w~\left|\hat{O}^{\Delta}_f\right|^2\right\}\exp\left\{\frac{f_i}{2}\int\!  \frac{d\Delta}{2\pi i}d^2w~|\hat{O}_i^\Delta|^2\right\}\right\rangle^{-1}\\
		&=G_{\Delta_f,\Delta_i}(w_f,\bar{w}_f;w_i,\bar{w}_i)\\
  &\hspace{1cm}+f_if_f\int\!\frac{d\Delta}{2\pi i}d^2w\celint{\Delta'}{w'}G_{\Delta_f,\Delta^*}(w_f,\bar{w}_f;w,\bar{w})G^{f_f,f_i}_{\Delta',\Delta}(w,\bar{w};w,\bar{w}')G_{\Delta'^*,\Delta_i}(w',\bar{w}';w_i,\bar{w}_i),
	\end{split}
\end{equation}
where, in defining the double deformed two-point function, we normalized the deformed vacuum.
Because in our case $G_{\Delta,\Delta'}(w,\bar{w};w',\bar{w}')=2\pi i \delta(\Delta-\Delta'^*)\delta^2(w-w')$, \AM{we obtain}
\begin{align}
	\hat{G}^{f_f,f_i}=2\pi i \delta(\Delta_i-\Delta^*_f)\delta(w_f-w_i)\frac{1}{1-f_f f_i}.
\end{align}
\paragraph{\AM{Hubbard-Stratonovich} method}
We can approximate this result in another way, called \AM{Hubbard-Stratonovich method }\cite{Hartman:2006dy}. We can deduce any $n$-point correlator in double deformed Celestial CFT which is dual to double deformed time-ordered $n$-point scattering in the bulk by varying the generating function 
\begin{equation}
    Z^{f_f,f_i}[J_f,J_i]:=\left\langle \exp\left\{\int\!d^2w\frac{d\Delta}{2\pi i}~ J^\Delta_f \hat{O}^\Delta_f+\frac{f_f}{2}\hat{O}_f^\Delta \hat{O}_f^{\Delta^*}\!\right\}\exp\left\{\int\!d^2w\frac{d\Delta}{2\pi i}~ J^\Delta_i \hat{O}^\Delta_i+\frac{f_i}{2}\hat{O}_i^\Delta \hat{O}_i^{\Delta^*}\!\right\} \right\rangle\label{eq:Hubbard}
\end{equation}
by $J_i,~J_f$. We take $f_i,f_f \in \mathbb{R}$.
We introduce an auxiliary field $\sigma$ as
\AM{\begin{equation}
\begin{split}
     &\left\langle \exp\left\{\int J_fO_f+\frac{f_f}{2}O_f^2\right\}\exp\left\{\int J_i\hat{O}_i+\frac{f_i}{2}\hat{O}_i^2\right\} \right\rangle\\
    &\propto\left\langle \exp\left\{\int J_f\hat{O}_f+\frac{f_f}{2}\hat{O}_f^2\right\}\int \mathcal{D}\sigma_f~ e^{-\frac{1}{2f_f}\int|\sigma_f+f_f\hat{O}_f|^2}\exp\left\{\int J_i\hat{O}_i+\frac{f_i}{2}\hat{O}_i^2\right\}\int\! \mathcal{D}\sigma_i ~e^{-\frac{1}{2f_i }\int|\sigma_i+f_i\hat{O}_i|^2} \right\rangle\\ 
   &= \left\langle\int\! \mathcal{D}\sigma_f~e^{\int \!d^2w\frac{d\Delta}{2\pi i}~ (J^\Delta_f-\sigma_f^{\Delta^*})\hat{O}^\Delta_f}\int \!\mathcal{D}\sigma_i~e^{\int \!d^2w\frac{d\Delta}{2\pi i}~ (J^\Delta_i-\sigma_i^{\Delta^*})\hat{O}^\Delta_i}~\exp{\left\{-\int \frac{|\sigma_f|^2}{2f_f}+\frac{|\sigma_i|^2}{2f_i}\right\}}\right\rangle\\
   &=\int \!\mathcal{D}\sigma_i\int \!\mathcal{D}\sigma_f~\exp\left\{\int\left(-\frac{|\sigma_f|^2}{2f_f}-\frac{|\sigma_i|^2}{2f_i}\right)+(J^\Delta_f-\sigma_f^{\Delta^*})\frac{G_{\Delta,\Delta'}}{2\pi i}(J^{\Delta'}_i-\sigma_i^{\Delta'^*})\right\}.\label{eq:HS method rough}
\end{split}
\end{equation}}
Substituting $G_{\Delta,\Delta'}(w,\bar{w};w',\bar{w}')=2\pi i\delta(\Delta-\Delta'^*)\delta^2(w-w')$, we have
\AM{\begin{equation}
    \begin{split}
        &\left\langle \exp\left\{\int J_f\hat{O}_f+\frac{f_f}{2}\hat{O}_f^2\right\}\exp\left\{\int J_i\hat{O}_i+\frac{f_i}{2}\hat{O}_i^2\right\} \right\rangle\\
        &\propto\int \!\mathcal{D}\sigma_i\int \!\mathcal{D}\sigma_f~\exp\int d^2w \frac{d\Delta}{2\pi i }\left\{-\frac{|\sigma_f|^2}{2f_f}-\frac{|\sigma_i|^2}{2f_i}+(J^\Delta_f-\sigma_f^{\Delta^*})(J^{\Delta^*}_i-\sigma_i^{\Delta})\right\}\\
        &\propto \exp\left[\frac{\int\!d^2w \frac{d\Delta}{2\pi i}~J^\Delta_f J^{\Delta^*}_i+f_fJ_f^\Delta J_f^{\Delta^*}/2+f_iJ_i^\Delta J_i^{\Delta^*}/2}{1-
        f_if_f}\right].
    \end{split}
\end{equation}}
This is the exact result in contrast to the approximate result by taking the large-$N$ limit in AdS/CFT case since we consider the free theory. Note that the divergence from vacuum bubble is removed since  $\sigma_f$ and $\sigma_i$ are commute. The two-point function $\langle \hat{O}_i^\Delta \hat{O}_i^\Delta \rangle,~\langle \hat{O}_f^\Delta \hat{O}_f^\Delta \rangle$ thereby differs from the one obtained from the direct calculation.

\subsection{non-local double deformation}\label{subsec:non-local}
Next, we calculate the two-point function of the non-local double deformed Celestial CFT.
Non-local \AM{double} deformation is represented as follows and can be calculated in the same way as the previous \AM{subsection,
\begin{equation}
    \begin{split}
        &\left\langle \hat{O}_f^{\Delta_f}:\exp\left\{{f}\int \frac{d\Delta}{2\pi i}~\hat{O}_f^\Delta \hat{O}_i^{\Delta^*}\right\}:\hat{O}_i^{\Delta_i}\right\rangle\\
        &=f\int\!d^2w \frac{d\Delta}{2\pi i}~G_{\Delta_f,\Delta}G_{\Delta^*,\Delta_i}\left\langle :e^{f\int \hat{O}_f^\Delta \hat{O}_i^{\Delta^*}}:\right\rangle+G_{\Delta_f,\Delta_i}\left\langle :e^{f\int \hat{O}_f^\Delta \hat{O}_i^{\Delta^*}}:\right\rangle+0\\
        &=f\int\!d^2w \frac{d\Delta}{2\pi i}~G_{\Delta_f,\Delta}G_{\Delta^*,\Delta_i}+G_{\Delta_f,\Delta_i},
    \end{split}
\end{equation}}
where $: \cdot :$ denotes the normal ordering. Substituting $G_{\Delta,\Delta'}(w,\bar{w};w',\bar{w}')=2\pi i\delta(\Delta-\Delta'^*)\delta^2(w-w')$, we have
\begin{equation}
    \begin{split}
         &\left\langle \hat{O}_f^{\Delta_f}:\exp\left\{f\int \frac{d\Delta}{2\pi i}~\hat{O}_f^\Delta \hat{O}_i^{\Delta^*}\right\}:\hat{O}_i^{\Delta_i}\right\rangle\\
         &=\left(1+f\right)(2\pi i)\delta(\Delta-\Delta'^*)\delta^2(w-w').
    \end{split}
\end{equation}
This is the two function under the non-local \AM{double deformation}\footnote{In this non-local \AM{double deformation,} using the CCFT operator formalism, we can evaluate massive two-point functions without difficulty unlike the local double-trace deformation case. We comments on the result little bit. By noting the undeformed celestial two-point function is given by
\begin{align}
   G_{\Delta',\Delta}(w',\bar{w}';w,\bar{w})= \frac{2\pi}{m^2(\Delta-1)}\frac{\delta(\Delta'-\Delta)}{|w-w'|^{2\Delta}}-\frac{2\pi^2}{m^2(\Delta-1)^2}\delta(\Delta-\Delta'^*)\delta^{(2)}(w-w'),
\end{align}
we can evaluate the two-point function under the non-locally \AM{double deformation}
\begin{equation}
    \begin{split}
        &\left\langle \hat{O}_f^{\Delta_f}:\exp\left\{-\frac{f}{2}\int d\Delta~\frac{|\Delta-1|^2}{32\pi^3 }\hat{O}_f^\Delta \hat{O}_i^{\Delta^*}\right\}:\hat{O}_i^{\Delta_i}\right\rangle\\
        &=-\frac{f}{2}\int\!d^2w d\Delta~\frac{|\Delta-1|^2 m^2}{32\pi^3 }G_{\Delta_f,\Delta}G_{\Delta^*,\Delta_i}+G_{\Delta_f,\Delta_i},
    \end{split}
\end{equation}
and we obtain
\begin{equation}
    \begin{split}
         &\left\langle \hat{O}_f^{\Delta_f}:\exp\left\{-\frac{f}{2}\int d\Delta~\frac{|\Delta-1|^2}{32\pi^3 }\hat{O}_f^\Delta \hat{O}_i^{\Delta^*}\right\}:\hat{O}_i^{\Delta_i}\right\rangle\\
         &=\left(\frac{2\pi}{m^2}-\frac{f}{4m^2}\right)\frac{1}{(\Delta_f-1)}\frac{\delta(\Delta_f-\Delta_i)}{|w-w'|^{2\Delta}}-\left(\frac{2\pi^2}{m^2}-\frac{5\pi^2f}{8m^2\pi}\right)\frac{1}{(\Delta_f-1)^2}\delta(\Delta_f-\Delta_i^*)\delta^{(2)}(w-w').
    \end{split}
\end{equation}
}.

This result can be interpreted in the bulk that the correlator \AM{becomes stronger }(weaker) if we insert the interaction $f>0(<0)$ between two asymptotic sides.


\section{AdS/CFT-like(Wedge-like) holography}\label{sec:Wedge-like hologrpahy}%
In this section, we compute ``Wedge dual" two-dimensional operator correlation function dual to a bulk massless scalar field using the GKPW relation. We show that the mixed boundary conditions, which are expected to dual to double-trace deformations, non-trivially change two-point functions. Also, we discuss the possibility that, by changing deformation parameters of mixed boundary conditions, whether we can reproduce celestial CFT two-point functions \eqref{eq:CCFTtwopointfunction}.

\subsection{Celestial operators and Wedge-like operators} \label{subsec:CCFTvsWCFT}
In Wedge holography, we apply AdS/CFT holography to EOW branes
in Minkowski space viewed as \AM{three-dimensional} (Euclidean) AdS space. This identification indicates that the Wedge operators live in the boundary \AM{limit }($\rho\to\infty$ in region $\mathcal{A}$ and $t\to\pm\infty$ limit in region $\mathcal{D}$) on EOW branes. Therefore, we firstly examine the behaviour of a bulk scalar field in the boundary limit of the EOW branes.
 \paragraph{$\rho\to\infty,~t\to\pm\infty$  limit}
For the region $\mathcal{D}$, a plane-wave solution can be expanded in the Mellin basis as 
\AM{\begin{equation}
	e^{\mp ikX_\mp}=\int^{1+i\infty}_{1-i\infty}\!\frac{d\Delta}{2\pi i} \phi^\mp_\Delta\omega^{-\Delta}=\int^{1+i\infty}_{1-i\infty}\!\frac{d\Delta}{2\pi i} \frac{(\pm i)^\Delta\Gamma(\Delta)\omega^{-\Delta}r^{-\Delta}}{(\sinh t-\cosh t\cos\tau\pm i\kappa')^\Delta}
\end{equation}}
by taking the inverse of Mellin transform \eqref{def:cpw}, using 
\begin{equation}
	-k(\omega,\Omega)\cdot X(r,t,\Omega_0)=\omega r(\sinh t-\cosh t\cos\tau(\Omega,\Omega_0)),
\end{equation}
\AMM{where $\cos\tau(\Omega,\Omega_0)$ is defined by
\begin{equation}
    \cos\tau(\Omega,\Omega_0) =\cos\theta\cos\theta_{0}+\sin\theta\sin\theta_{0}\cos(\varphi-\varphi_{0}).
\end{equation}
Below, we will denote $\tau(\Omega,\Omega')$ simply as $\tau$.}
In the limit $t\to\infty$, \AM{two modes} appear, a contact term and a non-contact term for each $\Delta$ as shown \AM{below }\cite{deBoer:2003vf}; 
\AMM{\begin{align}
		\left.\phi^\mp_\Delta\right|_{t\gg 1}&=\frac{\Gamma(\Delta)}{r^{\Delta}}\left[-\frac{4\pi}{\Delta-1}(\mp i)^{\Delta} e^{(\Delta-2)t}\delta^{2}(\Omega-\Omega_0)+(\pm i)^\Delta e^{-\Delta t}\left(\frac{2}{1-\cos\tau}\right)^{\Delta}\right],\\
	\left.\phi^\mp_\Delta\right|_{-t\gg 1}&=\frac{\Gamma(\Delta)}{r^{\Delta}}\left[-\frac{4\pi}{\Delta-1}(\pm i)^\Delta e^{(2-\Delta) t}\delta^{2}(\Omega-\Omega^*_0)+(\mp i)^\Delta e^{\Delta t}\left(\frac{2}{1+\cos\tau}\right)^{\Delta}\right],
\end{align}}
where we take $\Omega-\Omega_0^*=(\theta-\theta_0-\pi,\phi-\phi_0)$ in the second limit so that $\cos\tau=-1$.
Here, we need two conditions $\epsilon>0$ and $\kappa>0$ to ensure that the delta functions in the above expression \AM{are} well-defined\footnote{In deriving the above result, we used the following formula of the delta function
\begin{equation}
    \delta^{(2)}(x) = \frac{1}{\pi}\cdot \frac{\Gamma(\Delta)}{\Gamma(\Delta-1)}\lim_{\epsilon \to 0} \frac{\epsilon^{2\Delta-1}}{(x^{2}+\epsilon^{2})^{\Delta}}.
\end{equation}
To ensure the delta function correctly \AM{converges}, we need the condition $\epsilon>0$. 
Also, to deform the counter integral $\lim_{t\to\infty}\phi^\mp_\Delta$ and $\lim_{t\to -\infty}\phi^\mp_\Delta$ to be the form of the R.H.S. of the above equation, we further need condition $\kappa>0$.
See also appendix A of \cite{deBoer:2003vf} for the related discussion on the delta function here.}. 
Continuing analytically from the region $\mathcal D$ to the region $\mathcal A^+$, we obtain the expansion of the plane-wave in region $\mathcal A^+$. \AM{Here the analytic continuation} differs between in and out modes;
\begin{align}
	\mathcal{D}\to \mathcal{A}^+:&r\to i\eta,\quad t\to \rho-i\frac{\pi}{2}\quad \text{for} \quad e^{ikX_+},\\
	\mathcal{D}\to \mathcal{A}^+:&r\to -i\eta,\quad t\to \rho+i\frac{\pi}{2}\quad \text{for}\quad e^{-ikX_-}.
\end{align}
Then, in the Mellin basis, the plane-wave can be expanded as
\begin{align}
	e^{\mp ikX_\pm}=\int^{1+i\infty}_{1-i\infty}\!\frac{d\Delta}{2\pi i}~\omega^{-\Delta}\phi^\mp_\Delta= &\int^{1+i\infty}_{1-i\infty}\!\frac{d\Delta}{2\pi i}\frac{(\pm i)^\Delta \Gamma(\Delta)\omega^{-\Delta}(\eta\pm i\tilde\epsilon\eta)^{-\Delta}}{(\cosh\rho-\sinh\rho\cos\tau)^{\Delta}},
\end{align}
where $\epsilon$ and $\kappa'$ are related by
\begin{equation}
    \kappa'=(-q\cdot \hat{X})\tilde\epsilon,\quad \tilde\kappa>0
\end{equation}
since for region $\mathcal A^+$
\begin{equation}
	\cosh\rho-\sinh\rho\cos\tau>0.
\end{equation}
In the limit $\rho\to\infty$, 
\AMM{\begin{equation}
	\left.\phi^\mp_\Delta\right|_{\rho\gg 1}= \frac{(\pm i)^\Delta\Gamma[\Delta]}{\eta^\Delta_\pm}\left[\frac{4\pi}{\Delta-1}e^{(\Delta-2)\rho}\delta(\Omega-\Omega_0)+e^{-\Delta\rho}\left(\frac{2}{1-\cos\tau}\right)^\Delta\right].
\end{equation}}
\paragraph{Wedge-like operator} Now we are ready to  construct an Wedge-like operator dual to the Minkowski space. Recall that in AdS/CFT correspondence, we used the fact that there are two modes of a bulk solution, with the mode with conformal dimension $\Delta$ being dominant compared to the mode with conformal dimension $2-\Delta$. Therefore, we have to make a certain assumption on $\Delta$ in order to make use of our AdS/CFT techniques. \par 

Let us consider taking $\Re[\Delta]=1+\epsilon>1$. This regularization is reasonable if we consider removing the IR divergence. As we know, the conformal primary wavefunction \eqref{def:cpw} diverges for IR region $\omega\sim 0$. From the perspective of Celestial holography, this IR divergence relates to soft theorems for bulk gauge \AM{fields }\cite{Guevara:2019ypd, Puhm:2019zbl,Adamo:2019ipt,Pate:2019mfs,Fan:2019emx,Cheung:2016iub} in the way that singular points $\Delta=-m$ at which the residue can be captured by a celestial current $J\sim \lim_{\Delta\to-m}(\Delta+m)$ come from scattering behaviours $\mathcal{A}\sim \omega^m,~m=-1,0,1,\dots.$ 
On the other hand, from the point of Wedge-like view, we would like to regularize and remove the first contact term, which corresponds to $\Delta=1$ pole. Therefore, our procedure in this section has to be to insert $\epsilon>0$ first into $\Delta$ and take $\epsilon\to 0$ limit lastly as \cite{Costa:2014kfa} did to extract the delta-functional behaviour of the imaginary part of the conformal dimension. This procedure is  similar to the dimensional regularization when removing IR divergences\footnote{In usual AdS/CFT, we have the IR/UV relation, which implies that UV energy scales in CFT are related to IR ones in AdS. Since the Minkowski/CFT correspondence here uses the AdS/CFT, we can consider that the IR regularization in Minkowski corresponds to UV regularization in wedge CFT.}. In a more general bulk dimension\AM{, say, $(d+2)$-dimensional bulk spacetime}, the conformal dimension can be written as  $\Delta=\frac{d}{2}+i\nu$. Therefore, the dimensional regularization is equivalent to shifting $\Delta\to\Delta+\epsilon,~\epsilon>0$. Then we obtain a ``extrapolate dictionary"
\AMM{\begin{align}
	\left.\Psi(X)\right|_{\rho \gg 1}&= \int\frac{d\Delta}{2\pi i}~e^{(\Delta-2)\rho}\eta^{-\Delta}O_W^\Delta(\pm;\Omega) &\text{for \ra{\pm},}\\
	\left.\Psi(X)\right|_{t \gg 1}&= -\int\frac{d\Delta}{2\pi i}~e^{(\Delta-2)t}r^{-\Delta}O_W^\Delta(+;\Omega)\\
	\left.\Psi(X)\right|_{-t\gg 1}&= -\int\frac{d\Delta}{2\pi i}~e^{-t\Delta}r^{-\Delta}O_W^\Delta(-;\Omega)&\text{for \rd},
\end{align}}
where we defined \textit{Wedge-like operator} as
\begin{align}
		O^\Delta_W(\pm;\Omega)= {\Gamma[\Delta]}\frac{4\pi}{\Delta-1}\left[{ (\pm i)^\Delta} O^+_\Delta+{(\mp i)^\Delta} O^-_\Delta\right].
\end{align}
Note that ${O^\Delta_W}^\dagger=O_W^{2-\Delta}$. 
They are analytically continuous.
Unlike the $\kappa$, which was taken to converge the integral in the UV region, the sign $\Re[\Delta]-1$ does not change between the incoming wavefunction and the outgoing wavefunction. Thus, we can rewrite the bulk by the Wedge operator as
\AMM{\begin{align}
	&\left.\Psi_\Delta(X)\right|_{\rho \gg 1}&\notag\\
	&=e^{(\Delta-2)\rho}\int\! d^2\Omega_0\left[\delta(\Omega-\Omega_0)+e^{(2-2\Delta)\rho}\frac{\Delta-1}{4\pi}\left(\frac{2}{1-\cos\tau}\right)^\Delta \right]O_W^\Delta(\eta,\Omega_0)&\text{for region }\mathcal{A}\label{eq:OW expansion A+},\\
	&\left.\Psi_\Delta(X)\right|_{t \gg 1}&\notag\\
	&= \left[-e^{(\Delta-2)t}O_W^\Delta(+;r,\Omega)+e^{-\Delta t}\frac{\Delta-1}{4\pi}\int\! d^2\Omega_0\left(\frac{2}{1-\cos\tau}\right)^\Delta O_W^\Delta(-;r,\Omega_0)\right]&\text{for region }\mathcal{D}\label{eq:OW expansion D},\\
	&\left.\Psi_\Delta(X)\right|_{-t\gg 1}&\notag\\
	&= \left[-e^{(2-\Delta) t}O_W^\Delta(-;r,\Omega)+e^{\Delta t}\frac{\Delta-1}{4\pi}\int\! d^2\Omega_0~\left(\frac{2}{1+\cos\tau}\right)^\Delta O_W^\Delta(+;r,\Omega_0)\right]&\text{for region }\mathcal{D}\label{eq:OW expansion D-}.
\end{align}}

\AMM{


We used AdS/CFT techniques to obtain celestial two-point functions under mixed boundary conditions. In \cite{Iacobacci:2022yjo}, the authors also use AdS/CFT techniques to compute $n$-point celestial correlators.  It would be interesting to find the relation between ours and theirs by generalizing our computations to interacting cases.



}

\subsection{Changing boundary conditions on AdS/dS patch boundaries}\label{subsec:changingbdyCondi}

    In the AdS/CFT framework, transitioning from Neumann to Dirichlet boundary conditions in the bulk is a pivotal step for renormalizing the infrared divergent term in the bulk, which is called ``holographic renormalization". This is equivalent to removing a ultraviolet term in the dual CFT and therefore invokes a flow between conformal fixed points from UV to IR. We can also show that changing from Neumann to Dirichlet boundary conditions in the Minkowski space is also equivalent to a flow between different 2-codimensional wedge-like CFTs. This fact can be found by examining the two-point function as we well see in the end of this subsection. 
    To illustrate this, we first establish the GKPW dictionary of the Wedge-like holography and investigate the two-point function with mixed boundary conditions under the setup we explained in previous subsection \S\ref{subsec:CCFTvsWCFT}.
    Although the change in boundary conditions effectively removes a divergent term in codimension-two wedge holography, its implications for energy flow remain unclear. In AdS/CFT, the change of the boundary conditions and the resultant modification of the free energy is captured by a shift of the central charge \cite{Gubser:2002vv}.  However, we cannot apply the same discussion for asymptotic flat four-dimensional space, where central charge of the background graviton is both imaginary and divergent \cite{Pasterski:2022lsl,Ogawa:2022fhy}. This obscures its physical meaning. The detailed discussion is provided in Appendix \ref{appendix:cfunction}.

\paragraph{operator formalism and GKPW relation}
 We will prove the equivalence of an \AM{operator }(BDHM) formalism and GKPW dictionary with reference to the case of \AM{AdS/CFT }\cite{Terashima:2017gmc}.

  The action of a massless field with a cutoff at fixed $\rho=R,~t=t^+:=R,~t=t^-:=-R$ is 
\begin{equation}
	\begin{split}
		S[\gamma,\phi_b]=&S_{\mathcal A}+S_{\mathcal{D}}\\
		=&\frac{1}{2}\int\! d\rho d\eta d^2\Omega~\sqrt{-g}[\partial\Psi\partial\Psi]+\int_{\rho=R,\partial\mathcal{A}^++\partial\mathcal{A}^-} \hspace{-40pt}d\eta d^2\Omega ~\sqrt{-g}g^{\rho\rho}\left[\frac{1}{2}\gamma\Psi^2-\Psi\Psi_b\right]\\
		&+\frac{1}{2}\int\! dt dr d^2\Omega~\sqrt{-g}[\partial\Psi\partial\Psi]+\int _{t^+-t^-}\hspace{-20pt} dr d^2\Omega ~\sqrt{-g}g^{tt}\left[\frac{1}{2}\gamma\Psi^2-\Psi\Psi_b^\pm \right].
\end{split}
\end{equation}
The second and the forth term are boundary terms that are determined by boundary conditions as we will see soon. {The $\gamma$ terms are necessary for transitioning from Neumann boundary condition to Dirichlet boundary condition. }To see this,
let us consider the variation of the action with respect to $\Psi$,
\begin{equation}
\begin{split}
	\delta_\Psi S&=(EOM)\\
	&+\int_{\rho=R,\partial\mathcal{A}^++\partial\mathcal{A}^-} \hspace{-40pt}  d\eta d^2\Omega~\sqrt{-g}g^{\rho\rho}\delta\Psi[\partial_\rho\Psi+\gamma\Psi-\Psi_b]+\int_{t^+-t^-}\hspace{-20pt}  dr d^2\Omega~\sqrt{-g}g^{tt}\delta\Psi[\partial_t\Psi+\gamma\Psi-\Psi_b^\pm ],	\label{eq:wedge action}
	\end{split}
\end{equation}
where $\Psi(\eta,\Omega)$ and  $\Psi^\pm(r,\Omega)$ are analytically continued.
We first focus on region $\mathcal{A}$, and from \eqref{eq:wedge action}, the bulk equation of motion and the boundary condition are given by
\AM{\begin{align}
		\partial_\mu\partial^\mu\Psi(\eta,\rho,\Omega)&=0,\label{eq:massless EOM}\\
		\gamma\Psi(\eta,R,\Omega)+\partial_\rho\Psi(\eta,\rho,\Omega)|_{\rho=R}&=\Psi_b(\eta,\Omega).\label{eq:bdy cond. massless}
\end{align}}
The second condition is equivalent to mixing Neumann/Dirichlet boundary conditions on the boundary $\rho\to\infty$.
Similarly, from \eqref{eq:wedge action} the bulk equation of motion and the boundary condition in region $\mathcal{D}$ is
\AM{\begin{align}
\partial_\mu\partial^\mu\Psi(t,r,\Omega)&=0,\\
\gamma\Psi(t,r,\Omega)+\partial_t\Psi(t,r,\Omega)|_{t=\pm R}&=\Psi_b^\pm(r,\Omega). 	
\end{align}}

Noting that the free scalar field is satisfying the equation of motion, we can expand a free scalar as \eqref{eq:OW expansion A+}, \eqref{eq:OW expansion D} and \eqref{eq:OW expansion D-}
\begin{align}
		\Psi(\eta, R,\Omega)&=\int^{1+i\infty}_{1-i\infty} \!\frac{d\Delta}{2\pi i}d\Omega_0~e^{(\Delta-2) R}\left[\delta(\Omega-\Omega_0)+\frac{\Delta-1}{4\pi}e^{(2-2\Delta) R}\left(\frac{2}{1-\cos\tau}\right)^\Delta\right]O_W^\Delta(\Omega)\eta^{-\Delta}\notag\\
		&=:\int\!\frac{d\Delta}{2\pi i}  ~\hat{g}_\Delta O_W^{\Delta}(\Omega)\eta^{-\Delta}e^{(\Delta-2)R},\label{eq:ExpansionAdSPatch}\\
		\Psi(r, \pm R,\Omega)&=\int^{1+i\infty}_{1-i\infty} \!\frac{d\Delta}{2\pi i}d\Omega_0~e^{(\Delta-2) R}r^{-\Delta}\notag \\
		& \quad \times\left[-\delta(\Omega-\Omega_0)O_W^\Delta(\pm;\Omega)+\frac{\Delta-1}{4\pi}e^{(2-2\Delta) R}\left(\frac{2}{1\mp \cos\tau}\right)^\Delta O_W^\Delta(\mp;\Omega)\right]\notag\\
		&=:\int\!\frac{d\Delta}{2\pi i}  ~\hat{g}_\Delta O_W^{\Delta}(\Omega)\eta^{-\Delta}e^{(\Delta-2)R},\label{eq:ExpansiondeSitterPatch}
\end{align}
where we introduced a shorthand notation $\hat{g}_{\Delta}$. Then, from this expansion, we acquire a boundary field expansion
\AM{\begin{equation}
	\begin{split}
		\Psi_b=\int\!\frac{d\Delta}{2\pi i}e^{(\Delta-2)R}\eta^{-\Delta}\Psi_b^\Delta,\quad \Psi_b=\int\!\frac{d\Delta}{2\pi i}e^{(\Delta-2)R}r^{-\Delta}\Psi_b^{\Delta;\pm},
	\end{split}
\end{equation}}
where these fields are related to \eqref{eq:ExpansionAdSPatch} and \eqref{eq:ExpansiondeSitterPatch} as
\begin{equation}
	(\partial_\rho+\gamma)[\hat{g}_\Delta O^\Delta_W e^{(\Delta-2)\rho}]|_{\rho=R}=\Psi_b^\Delta,\label{eq:relation of phib and OW}
\end{equation}
\AM{\begin{equation}
    (\partial_t+\gamma)[\hat{g}_\Delta O^\Delta_W e^{(\Delta-2)t}]|_{t=\pm R}=\Psi_b^{\Delta;\pm}.\label{eq:relation of phib and OW for dS}
\end{equation}}
Let us focus on the region $\mathcal{A}^+$. Using these, to the leading order the left hand side of equation \eqref{GKPW} becomes 
\begin{equation}
	\begin{split}
		&\delta_{\Psi^\Delta_b}\log\text{(L.H.S.)}\\\
		&=\int\mathcal{D}\Psi_b ~e^{i\int {d\eta} d^2\Omega~\eta e^{2 R}\Psi(\eta, R,\Omega)\delta\Psi_b(\eta,\Omega;R)}/\int\mathcal{D}\Psi_b~e^{iS}\\
		&= \frac{1}{\int\mathcal{D}O_W~e^{iS}} \int\mathcal{D}{O_W}\\
  &\times\exp\left[\int\! {d\eta}\eta d^2\Omega~e^{2 R}\left(\int\!\frac{d\nu}{2\pi }  ~\hat{g}_\nu O^W_{\nu}(\Omega)\eta^{-\Delta}e^{(i\nu-1+\epsilon)R}\right)\left(\int\!\frac{d\nu'}{2\pi}e^{(i\nu'-1+\epsilon)R}\delta\Psi_b^{\nu'}\eta^{-\Delta'}\right)\right]\\
		&\simeq\int\mathcal{D}{O_W}\exp\left[ e^{2\epsilon R}\!\int\! d^2\Omega\int\!\frac{d\nu}{2\pi }~O_W^{\nu}(\Omega)\delta\Psi_b^{-\nu}(\Omega) \right]/\langle 1\rangle_{\text{CFT}}.
	\end{split}
\end{equation}
Therefore, we conclude that the GKPW relation is \eqref{GKPW} and the two-point function is calculated as 
\begin{equation}
	\begin{split}
		\langle \hat{O}_W^\Delta(\Omega) \hat{O}_W^{\Delta'}(\Omega')\rangle=\left.(2\pi i)^2e^{-4\epsilon R}\frac{\delta^2 }{\delta\Psi_b^{\Delta^*}(\Omega) \delta\Psi_b^{\Delta'^*}(\Omega')} \right|_{\Psi_b=0}\hspace{-20pt}S.
	\end{split}
\end{equation}

\paragraph{two-point functionl}\label{wedge- two-point function}
 However, instead of supposing \AM{the} na\"ive GKPW correspondence\footnote{More accurately, we have to care about the subleading order of the GKPW dictionary. 
 Since the leading term  of the $\Psi_b$ in $e^{2\epsilon\rho}$ vanishes when we consider holographic renormalization,
 there is no problem looking only at the leading term of the GKPW relation for the \AM{CFT }(IR) fixed point. On the other hand, the two-point function becomes different in middle points of RG flow from the GKPW.}, we revisit how Wedge-like CFT relates to the bulk partition function to obtain the two-point function. \par 
 
By performing the path integral of the bulk partition function, i.e. substituting the solution of the equation of motion $ \Psi=\int \hat{g}_\Delta O_\Delta  $ into the bulk partition function, this codimension-2 CFT can be realized as\footnote{We can also consider another approach to evaluate the bulk partition function. In evaluating the partition function \eqref{eq:bulk to wedge, strictly}, first we do the $\eta$-integral, but in principle we can leave the $\eta$-integral until the end of the calculation. The latter procedure would be suitable for cases that we consider a finite wedge region bounded by $ 0< \eta_{1}<  \eta_{2}< \infty $. However, for that case, it is difficult to evaluate the partition function further, thus we mainly focus on the first case by taking the limit $\eta_{1}\to 0$, $\eta_{2}\to \infty$.  }
\begin{equation}
	\begin{split}
		&Z^{\mathcal{A}^+}_\ft[\Psi|_{\text{bdy cond. }\phi_b}]/Z^{\mathcal{A}^+}_\ft[\Psi|_{\phi_b=0}]\\
		&=\int\mathcal{D}O_W ~e^{i\int_0^\infty {d\eta} d^2\Omega~\eta e^{2 R}\Psi(\eta, R,\Omega)\Psi_b(\eta,\Omega)}\\
		&=\int\mathcal{D}{O_W}\exp\left[\int\! {d\eta}\eta d^2\Omega~e^{2 R}\left(\int\!\frac{d\Delta}{2\pi i}  ~\hat{g}_\Delta O_W^{\Delta}(\eta,\Omega)e^{(\Delta-2)R}\right)(\partial_\rho+\gamma)\left(\int\!\frac{d\Delta'}{2\pi i}e^{(\Delta'-2)R}\hat{g}_{\Delta'}O_W^{\Delta'}\right)\right]\\
		&=\int\mathcal{D}{O_W}\exp\left[\!\int\!d\eta \eta d^2\Omega~e^{2R}\frac{d\Delta}{2\pi i}\frac{d\Delta'}{2\pi i}~(\eta^{-\Delta}e^{(\Delta-2)R}\hat{g}_\Delta O_W^{\Delta})(\Omega)(\partial+\gamma)(\eta^{-\Delta'}e^{(\Delta'-2)R}\hat{g}_\Delta O_W^{\Delta'})(\Omega) \right]\\
        &=\int\mathcal{D}{O_W}\exp\left[\!\int\! d^2\Omega\frac{d\nu}{2\pi}~e^{2R}(e^{(i\nu-1+\epsilon) R}\hat{g}_\nu O_W^{\nu})(\Omega)(\partial+\gamma)(e^{(-i\nu-1+\epsilon) R}\hat{g}_{-\nu} O_W^{-\nu})(\Omega) \right].\label{eq:bulk to wedge, strictly}
	\end{split}
\end{equation}

Here, let us focus on region $\mathcal{A}^+$. Then, the above integrand can be calculated as
\begin{equation}
\begin{split}
	e^{2R}&(e^{(i\nu-1+\epsilon) R}\hat{g}_\nu O_W^{\nu})(\partial+\gamma)(e^{(-i\nu-1+\epsilon) R}\hat{g}_{-\nu} O_W^{-\nu})[\Omega,\Omega_0']\\
	&=e^{2 R}\left(\int d\Omega_0\left[e^{(i\nu-1+\epsilon) R}\delta(\Omega-\Omega_0)+\frac{i\nu}{4\pi}e^{(-i\nu-1-\epsilon) R}\left(\frac{2}{1-\cos\tau}\right)^{1+i\nu}\right]\right.\\
	&\hspace{20pt}\times(\partial+\gamma)\left.\int d\Omega'_0\left[e^{(-i\nu-1+\epsilon) R}\delta(\Omega_0-\Omega'_0)-\frac{i\nu}{4\pi}e^{(i\nu-1-\epsilon) R}\left(\frac{2}{1-\cos\tau'}\right)^{1-i\nu}\right]\right)\\
	&=\left[(\gamma-i\nu-1+\epsilon)e^{2\epsilon R}\delta(\Omega-\Omega')\right.\\
 &\qquad +e^{-2\epsilon R}(\gamma+i\nu-1-\epsilon)\frac{i\nu}{4\pi}\frac{-i\nu}{4\pi}\int d\Omega_0 \left(\frac{2}{1-\cos\tau}\right)^{1+i\nu}\!\left(\frac{2}{1-\cos\tau'}\right)^{1-i\nu}\\
	&\left.\quad +(\gamma+i\nu-1-\epsilon)e^{2i\nu R}\frac{-i\nu}{4\pi}\left(\frac{2}{1-\cos\tau}\right)^{1-i\nu}+\!(\gamma-i\nu-1+\epsilon)e^{-2i\nu R}\frac{i\nu}{4\pi}\left(\frac{2}{1-\cos\tau}\right)^{1+i\nu}\right].
\end{split}
	\end{equation}

 In Wedge holography, we restrict the dual bulk to the region surrounded by EOW branes within $\mathcal{A}^+$.  Therefore, it is meaningful to see the two-point function made by the restricted  action $S_{\mathcal{A}^+}$ to the region $\mathcal{A}$. The action is
 \begin{equation}
 	S_{\mathcal{A}^+}\simeq\int d^2\Omega d^2\Omega_0\frac{d\Delta}{2\pi i}\left[(\gamma-1+\epsilon)\delta(\Omega-\Omega)e^{2\epsilon R}+(\gamma-i\nu-1)e^{-2i\nu R}\frac{\Delta-1}{4\pi}\left(\frac{2}{1-\cos\tau}\right)^\Delta\right]O_W^{\Delta}O_W^{2-\Delta}.\label{eq:MasslessActionOnAdSPatch}
 \end{equation}
  By taking $\gamma=1$, which amounts to the holographic renormalization, we have
\begin{equation}\label{eq:twogamma1}
	\begin{split}
	\langle {\hat{O}}_W^\Delta{\hat{O}}_W^{\Delta'}\rangle\propto e^{i(\Delta-1)R}\frac{1}{8\pi}\left(\frac{2}{1-\cos\tau}\right)^{2-\Delta}\delta(\Delta+\Delta'-2).
	\end{split}
\end{equation}

Incidentally, we can compare this result with the (massless) two-point function derived from Celestial CFT;
\AM{\begin{equation}
\begin{split}
    	\langle \hat{O}^\Delta_W \hat{O}^{\Delta'}_W\rangle &=-{\Gamma[\Delta]}\Gamma[\Delta']\left(\frac{4\pi}{\Delta-1}\right)\left(\frac{4\pi}{\Delta'-1}\right)\langle\left[{ i^\Delta}O^+_\Delta+{(- i)^\Delta} O^-_\Delta\right]\left[{ i^{\Delta'}}O^+_{\Delta'}+{(- i)^{\Delta'}} O^-_{\Delta'}\right]\rangle\\
    	&=-2\pi i~{\Gamma[\Delta-1]\Gamma[1-\Delta]}(4\pi)^2\delta(\Delta+\Delta'-2)\delta(\Omega-\Omega')\label{two-point fn:cel}\\
     & \propto \delta(\Delta+\Delta'-2)\delta(\Omega-\Omega'). 
\end{split}
\end{equation}}
Now we can explain why this difference between Wedge-like holography's result and Celestial Holography's result occurs. We took $\epsilon>0$ to make use of the AdS/CFT formulation and renormalize so that wedge dual \AM{two}-dimensional theory becomes CFT. In Celestial holography, we \AM{do not} have to renormalize. We just take $\epsilon\to 0$ limit.  
In other words, from that AdS/CFT intuition that bulk mixed boundary conditions correspond to double trace deformation of \AM{the} dual CFT, the Wedge CFT fixed point might not coincide with Celestial CFT fixed point under the double-traced flow of our CFT dual to the bulk massless scalar \AM{field.}
Indeed, under this operation, the action becomes
\begin{equation}
    \begin{split}
        S[O_W]\simeq\int d^2\Omega\frac{d\nu}{2\pi}\left[2(\gamma-1)\delta(\Omega-\Omega')+2(\gamma-1-i\nu)e^{-2i\nu R}\frac{i\nu}{4\pi}\left(\frac{2}{1-\cos\tau}\right)^{1+i\nu}\right]O_W^\nu O_W^{-\nu}
    \end{split}.\label{eq:BetterAction}
\end{equation}
Here, by taking $R\to\infty$, the second term vanishes due to the Riemann-Lebesgue lemma. Thus the two-point function becomes
\AM{\begin{equation}
\begin{split}
    	\langle \hat{O}^\Delta_W \hat{O}^{\Delta'}_W\rangle =2\pi i\frac{1}{1-\gamma}\delta(\Delta+\Delta'-2)\delta(\Omega-\Omega'),
  \end{split}
\end{equation}}
which matches \eqref{two-point fn:cel} up to the overall factor.
{
Our analysis in this subsection has revealed that a parameter $\gamma$, which is responsible for changing the boundary conditions in the bulk, acts as a holographic renormalization factor and as a parameter which moves between different CFT fixed points in the context of Wedge-like holograhy. This behaviour is very similar to the case of AdS/CFT. In contrast, we cannot uncover the role of $\gamma$ from our analysis. A clearer understanding might be achieved by examining a massive case, since we have both contact and non-contact terms for massive case.
}\par 
So far, we have focused only on the $\mathcal{A}^{+}$. However, we can extend the above analysis to regions beyond wedge regions. We are currently analyzing such situations to construct the holography beyond wedge regions which we call ``AdS/CFT-like holography" in a broader sense. We anticipate sharing the outcomes of our research in the near future.

\subsection{Changing boundary conditions on EOW branes: Wedge holography} \label{subsec: Wedge-like holography}

In this subsection, we derive a massless bulk scalar two-point function with boundary conditions not on AdS/dS patch boundary but on EOW branes.
Using these boundary conditions, we can obtain a condition on the allowed conformal dimensions, and solve the condition analytically. It exactly leads to the analytic expression of discretized conformal dimensions. We can compare this analytical result with a similar discussion in \cite{Ogawa:2022fhy} for a massive scalar case. The authors obtained a similar condition on conformal dimensions, for the massive scalar case, one can not analytically solve the condition on conformal dimensions, and they solved the condition numerically.

We focus on the region $\mathcal{A}^{+}$. In this case, a massless bulk scalar action with boundary terms is given by
\begin{equation}
	\begin{split}
        S_{\mathcal{A}}=&\int d^2\Omega d\eta d\rho~\eta^3\sinh^2\rho\frac{1}{2}\partial\Psi(\eta,\rho,\Omega)\partial\Psi(\eta,\rho,\Omega)\\
        & \qquad - \int^{R}_0d\rho d^2\Omega~\frac{f}{2}\sinh^2\rho[\eta_2^2\Psi^2(\eta_2)-\eta_1^2\Psi^2(\eta_1)]\\
        &\hspace{4cm}  +\int_{\rho=R} d^2\Omega ~e^{2R}\int^{\eta_2}_{\eta_1} d\eta \Psi(\rho=R) \Psi_b (\eta,\rho=R,\Omega),
	\end{split}
\end{equation}
where $\rho=R \gg 1$ is the cut off for $\rho$-direction, and $\eta_{1,2}>0$ labels the locations of EOW branes.
Varying the action with respect to $\Psi$, we obtain the \AM{equations} of motion for the bulk field $\Psi$ and the boundary conditions
\begin{align}
    [\eta\partial_\eta\Psi(\eta,\rho,\Omega)+f\Psi(\eta,\rho,\Omega)] |_{\eta=\eta_{1},\eta_{2}}&=0\label{bdy cond:eta},\\
    [\partial_\rho\Psi(\eta,\rho,\Omega)- \Psi_b(\eta,\rho,\Omega)]|_{\rho=R}=0.
\end{align}
From the boundary condition \eqref{bdy cond:eta}, 
as is the discussion in \ref{subsec:changingbdyCondi}, we can expand the scalar field as 
\begin{align}
		\Psi(\eta, R,\Omega)&= \sumint_{\Delta} d\Omega_0~e^{(\Delta-2) R}\left[\delta(\Omega-\Omega_0)+\frac{\Delta-1}{4\pi}e^{(2-2\Delta) R}\left(\frac{2}{1-\cos\tau}\right)^\Delta\right]O_W^\Delta(\Omega)\eta^{-\Delta}\notag,
\end{align}
where, unlike the case in subsection \ref{subsec:changingbdyCondi}, in the above expansion, mode $\Delta$ appearing in the integral (or sum) $\sumint_{\Delta} $ is not specified but still formal one here. To satisfy  $\eta$-boundary conditions at $\eta_{1,2}$ respectively, the condition $\eta_{1}^{\Delta-1}=\eta_{2}^{\Delta-1}$ is required.  This implies that $\Delta$ must be in the form
\begin{equation}
	\Delta=1+i\frac{2\pi n}{\log\left[\frac{\eta_1}{\eta_2}\right]} \qquad(n\in \mathbb{Z}),
\end{equation}
that is, $\Delta$ must be discretized unlike the case in subsection \ref{subsec:changingbdyCondi}. Combining this result and the discussion used in \ref{subsec:changingbdyCondi}, we can check that  the two-point function does not change for conformal dimensions in the form $\Delta=1+i\frac{2\pi n}{\log\left[\frac{\eta_1}{\eta_2}\right]}$\footnote{Precisely, we need to replace delta functions with Kronecker deltas, for example $\delta(\Delta+\Delta'+2) \to  \delta_{\Delta,2-\Delta}$.},   and  two-point function for other conformal dimensions are  identically zero: $\langle O_\Delta O_{\Delta'}\rangle=0$.

\section{Discussions and future directions}\label{sec:conclusion}

\paragraph{Summary and discussions}

In the celestial holography approach, we found that the deformed asymptotic conditions we have studied in this paper correspond to double deformation in celestial \AM{CFT,} which is similar to the double-trace-like deformation in AdS/CFT.

In the Wedge-like holography approach, we imposed the mixed Neumann/Dirichlet boundary conditions and computed bulk two-point functions.  We found that, from the bulk two-point function behaviour, the mixed boundary conditions in bulk trigger an RG flow on the dual CFT side, from renormalized IR fixed point, which corresponds the Wedge CFT, to UV fixed point, which corresponds to the Celestial CFT, if we look just at a massless scalar. 

We give some \AM{discussions} below;
\begin{itemize}
    \item \textit{Two types of double deformation in CCFT;} In celestial holography, since there are initial and final boundaries, we can naturally construct two types of double deformation. This is an important point of our double deformation.
    We can compare these double deformations with the original double trace deformations in the usual AdS/CFT correspondence. For pure AdS and one-sided AdS black holes, which have one boundary, there is a single type of double trace deformations. On the other hand, since the two-sided eternal AdS black hole has two distinct AdS boundaries, we can consider two types of double trace deformations.
    
    In the above sense, our double deformations in celestial holography are close to the original double trace deformations in AdS/CFT for two-sided AdS black holes.

        \item \textit{C-function under the change of boundary condition;} We have seen the RG-flow behavior of wedge holography by observing the two-point function in subsection \ref{subsec:changingbdyCondi}. 
        In wedge-like holography, we derived the GKPW relation from the BDHM dictionary, and we obtained the partition function. 
        One can also focus on a c-function of wedge holography to see the RG flow. 
         We can derive the c-function in the same way as AdS/CFT \cite{Gubser:2002vv,Hartman:2006dy}. In appendix \ref{appendix:cfunction}, we will explain how to evaluate the c-function  analytically under a certain limit. The resulting c-function is given by \eqref{eq:c-function}.
         If we look only at each $\Delta$, the resulting RG-flow behaviour seen from c-function of wedge holography is parallel to the behaviour in the \AM{AdS/CFT }\cite{Gubser:2002vv}. This observation for c-function is consistent with that for two-point functions in subsection \ref{subsec:changingbdyCondi}. However, the total change from the CFT fixed point of the c-function for all $\Delta$ vanishes. 
        %
        Summing up all modes $\Delta$ in boundary side  corresponds to accumulating all the AdS slices to make the Minkowski spacetime in bulk side. It would be interesting to find the bulk interpretation of the vanishing total c-function.


    \item \textit{Solving boundary conditions using Harmonic basis;} In wedge-like holography, we solve the boundary conditions  \eqref{eq:bdy cond. massless} as discussed in the body of this paper, and using the result, we evaluated two-point functions. However, we can adopt another way to solve the boundary conditions and evaluate two-point functions; the harmonic basis approach. We explain the method in appendix \ref{appendix:harmonic basis}. 
    In the appendix, we consider the case for region $\mathcal{A}$, but we \AM{can} do similar analyses for $\mathcal{D}$. 
    
    The merit of the harmonic basis approach is that two-point functions in the harmonics basis are written in a form such that one can compare the two-point functions with ones obtained in AdS/CFT. Indeed, two-point functions in the harmonics basis have a similar form as those derived in AdS/CFT, e.g., \cite{Hartman:2006dy}. In \cite{Hartman:2006dy}, the authors show that bulk two-point function under mixed boundary conditions coincides with CFT two-point function under corresponding double trace deformation. Thus, the above similarity of two-point functions between our two-point functions and AdS/CFT ones supports our claim that the mixed boundary conditions in bulk trigger an RG flow on the dual CFT side in the wedge-like holography.

\end{itemize}

\paragraph{Future directions}
 
We conclude this paper by listing several future directions.

\begin{itemize}
  
   


      \item \text{Deviation from the usual CCFT results}: In section \ref{sec:Wedge-like hologrpahy}, we computed the two-point functions by using the Wedge-like holography.
    \AM{However,} the obtained two-point functions are a little bit different from the known results of celestial CFT. This might be because there would be an essential difference between celestial holography and the Wedge-like holography.
    It would be interesting to investigate the difference and relations between them deeply. 

    \item \textit{Exotic Boundary Condition Related to Celestial Holography?} (work in progress): When we consider a massless scalar field with imposing the exotic boundary conditions as in \ref{appendix:holograEOW}, we can obtain two-point functions, which seems to be consistent with an expected celestial CFT two-point function.
    It would be interesting to study why such boundary conditions lead to the expected result consistent with celestial CFT. 
    
    \item \textit{Massive case?} (work in progress): In this paper, we mainly focused on bulk massless scalar fields for simplicity, but we can do almost similar analyses for bulk massive scalar fields as  briefly discussed in appendix \ref{appendix:massiveScalar}. Moreover, our disscussion was restricted to $\Im[\Delta]=\nu\geq0$, from the spanning a scalar of the dual bulk on the principal continuous series. If we look at more wider Celestial CFT including negative $\nu$, the massive two-point function behaves like middle point of RG flow in the sense of AdS/CFT. Therefore, we might extract richer physical meanings if we investigate more on massive \AM{case \cite{WIP:massivedd}.}
    
    

    \item \textit{Interacting case?}: Although we have considered non-interacting theories to simplify the discussions, we expect that even when there are interactions, we would be able to perturbatively treat double-trace deformations as in usual AdS/CFT.

    \item \textit{Non-Abelian and linearized gravitational theories?}: It would be interesting to consider our formulation of celestial holography, and the double-trace(-like) deformations for Non-abelian gauge theories and also the linearized gravitational theories. 

    \item \textit{$T\bar{T}$ deformation and $J\bar{T}$ deformation?}: It would be interesting to relate our discussions with $T\bar{T}$ and $J\bar{T}$ deformations in celestial holography e.g., \cite{He:2022zcf}.

    \item \textit{RG flow interpretation?}: As we commented above, we found that the changing asymptotic condition in specific ways induces the double deformation in the celestial CFT. However, we do not have a clear understanding of the physical interpretation of the double deformation itself within the context of CCFT. 
\end{itemize}

\section*{Acknowledgement}

We thank Yoshio Kikukawa for useful discussions in the early stage of this project and for useful comments on the draft. MF thanks Sotaro Sugishita, Hayato Hirai for useful discussions in the related projects.  MF also thanks to Hoiki Liu for correspondence. AM thanks Yuki Matsumoto for useful discussions in the related projects. AM also thanks Tadashi Takayanagi for correspondence.


 \appendix
 \section{Path integrals in coherent basis}\label{appendix:coherent}
 In this appendix, we explain useful path integral formulation for constructing Celestial holography, coherent path integral.
 
  \AMM{For notational convenience, in this appendix, we use $t$ to represent the time coordinate as in section \ref{sec:celestial dd}.}
\subsection{coherent states and some related formulae}
 In Heisenberg picture, scalar fields are generally written as
\AM{ \begin{align}
     \Psi(X)&=\int\! \frac{d^3\bk}{2\omega_\bk}~\left[a(\bk,t)e^{ik\cdot {X}}+a^\dagger(\bk,t)e^{-ik \cdot {X}}\right].
 \end{align}}
In particular, asymptotic fields which are defined by free fields satisfying
\begin{align}
    \lim_{t\to-\infty}[\Phi(\mathbf{x},t)-\Phi_\tin(\mathbf{x},t)]=0,\quad \lim_{t\to\infty}[\Phi(\mathbf{x},t)-\Phi_\tout(\mathbf{x},t)]=0
\end{align}
are written as
\AM{\begin{align}
    \Psi_\tin(X)&=\int\! \frac{d^3\bk}{2\omega_\bk}~\left[a_\tin(\bk)e^{ik\cdot {X}}+a^\dagger_\tin(\bk)e^{-ik \cdot {X}}\right],\\
    \Psi_\tout(X)&=\int\! \frac{d^3\bk}{2\omega_\bk}~\left[a_\tout(\bk)e^{ik\cdot {X}}+a_\tout^\dagger(\bk)e^{-ik \cdot {X}}\right],
\end{align}}
where
\begin{align}
    a_\tin(\bk)=\lim_{t\to -\infty}a(\bk,t),\quad a_\tout(\bk)=\lim_{t\to \infty}a(\bk,t).
\end{align}
		Assuming scalar fields asymptotically free, the annihilation and creation operators at asymptotic  infinity ($a_\tin/a_\tout$) and at finite time under  unitary time evolutions satisfy
		\AM{ \begin{align}
	 		[\hat{a}_\tin(\mathbf{k}),\hat{a}_\tin^\dagger(\bk)]=2\omega_\bk\dt{\bk'-\bk},&\quad [\ha_\tin(\bk),\ha_\tin(\bk')]=0,\quad  [\hat{a}_\tin^\dagger(\mathbf{k}),\hat{a}_\tin^\dagger(\bk')]=0,	\\
	 		[\ha(\bk,t),\ha^\dagger(\bk,t)]&=2\omega_\bk\dt{\bk'-\bk}.
	 	 \end{align}}
 A vacuum state in asymptotic states is defined to vanish by annihilation operators;
	\begin{align}
		\ha_\tin(\bk)|0\rangle_\tin=0,\quad {}_\tout \langle 0|\ha^\dagger_\tout(\bk)=0.
	\end{align}
Coherent states are defined as eigenstates of annihilation operators
\AMM{\begin{align}
	\ha (\bk,t)|z\rangle=z(\bk,t)|z\rangle,\quad \langle z |\ha^\dagger(\bk,t)=\langle z|\bar{z}(\bk,t).
\end{align}}
For free fields, we can construct coherent states explicitly
\begin{align}
   |z(\bk,t)\rangle&:=e^{-\frac{1}{2}\int d\tilde{k}|z(\bk,t)|^2} e^{\int d\tk~z(\bk,t)\hat{a}^\dagger(\bk,t)}|0\rangle. 
\end{align}
%
These coherent states \AM{form} the over-complete basis of the Hilbert space at each time.
We see this fact explicitly for asymptotic Hilbert space. These states obey
\AM{\begin{align}
		\langle z(\bk)|\beta(\bk)\rangle&=e^{-\frac{1}{2}\int d\tk(|z(\bk)|^2+|\beta(\bk)|^2)}\langle0|e^{\int d\tk~z^*(\bk)\hat{a}(\bk)}e^{\int d\tk~\beta(\bk)\hat{a}^\dagger(\bk)}|0\rangle
		\notag\\
  &=e^{-\frac{1}{2}\int d\tk(|z(\bk)|^2+|\beta(\bk)|^2)}e^{\int d\tk~z^*(\bk)\beta(\bk)}.
\end{align}}
When we take $\beta=z$, we can see that these states are normalized via
\AM{\begin{align}
	\langle z(\bk)|z(\bk)\rangle=1.
\end{align}}

\AMM{We can also show that the resolution of identity for on-shell is given by}
\AM{\begin{align}
	\hat{I}=\int \frac{\mathcal{D}z \mathcal{D}\bar{z}}{ V}|z\rangle\langle z|,
\end{align}
where $V$ is given by
\begin{align}
	V=\prod_{\bk}(2\pi i)2\omega_\bk.
\end{align}}
proof : Let $|n\rangle:=\int d\bk ~f(\bk)|n(\bk)\rangle$ be a superposition of $n$-particle Fock states. 
\AM{\begin{equation}
	\begin{split}
		\int \frac{\mathcal{D}z \mathcal{D}z}{ V}\langle n|z\rangle\langle z|m\rangle&=\int \frac{\mathcal{D}z \mathcal{D}z}{V}\int d\bk~f(\bk)z^n(\bk)\int d\bk ~f'(\bk')\bar{z}^m(\bk')\langle 0|z\rangle\langle z|0\rangle\\
		&=\int \frac{\mathcal{D}z \mathcal{D}z}{ V}\int d\bk ~f(\bk)z^n(\bk)\int d\bk ~f'(\bk')\bar{z}^m(\bk')e^{-\int d\tilde{k}''|z(\bk'')|^2}\\
		&=\prod_{\bk}\int \frac{dz(\bk) d\bar{z}(\bk)}{2\pi i(2\omega_\bk)} e^{-\frac{|\al{\bk}|^2}{2\omega_{\bk} }}z^n(\bk)\bar{z}^m(\bk)f(\bk)f'(\bk).\\
\end{split}
\end{equation}}
Replacing $z(\bk)$ by $\sqrt{2\omega_\bk (2\pi)^3}\rho_\bk e^{i\varphi_\bk}$, 
\AM{\begin{align}
	\begin{split}
		&\int \frac{dz(\bk) d\bar{z}(\bk)}{2\pi i(2\omega_\bk)} e^{-\frac{|\al{\bk}|^2}{2\omega_{\bk}}}z^n(\bk)\bar{z}^m(\bk)f(\bk)f'(\bk)\\&=2 \int^\infty_0 \rho_{\bk}d\rho_{\bk}\int^{2\pi}_{0} \frac{d\varphi_{\bk}}{2\pi}~e^{-{\rho^2_{\bk}}}\sqrt{(2\omega_\bk)^{(n+m)}}\rho_\bk^{n+m}e^{i\varphi_{\bk}(n-m)}f(\bk)f'(\bk)\\
	&=2\sqrt{(2\omega_\bk)^{(n+m)}}\int^\infty_0d\rho_\bk~\rho_\bk^{n+m+1}\delta_{n,m}e^{-\rho_\bk^2}f(\bk)f'(\bk)\\
	&=2(2\omega_\bk)^{n} \delta_{n,m}\int^\infty_0 d\rho_\bk~\rho_\bk^{2n+1}e^{-\rho^2_\bk}f(\bk)f'(\bk)\\
	&=(2\omega_\bk)^{n}\delta_{n,m}\int^\infty_0 dx~x^ne^{-x}f(\bk)f(\bk')\\
	&=n!\times (2\omega_\bk)^n \delta_{n,m}f(\bk)f(\bk')\\
	&=\langle n(\bk)|m(\bk)\rangle f(\bk)f(\bk').
	\end{split}
	\end{align}} 
 \qed

\subsection{Derivation of coherent path integral} We derive coherent path integral formulations.  
%
%

\paragraph{Derivation of coherent path integrals}
A coherent path integral is derived by inserting the resolution of identity 
\AM{\begin{equation}
	\begin{split}
		_{\text{H}}\langle f,t_f|i,t_i\rangle&_{\text{H}}\left(={}_{\text{S}}\langle f| e^{-i\int dt~\hat{H}(t)}|i\rangle_{\text{S}}\right)\\
		&=\langle f|\left(\prod_I\int\frac{\mathcal{D}z(t_I)\mathcal{D}\bar{z}(t_I)}{ V}|z(t_I)\rangle\langle z(t_I)|\right)|i\rangle,
	\end{split}
\end{equation}}
where $H$ denotes the state in Heisenberg picture and $S$ denotes the state in Schr\"odinger picture. The overlapping of coherent states with slight time difference is
\AM{\begin{equation}
	\begin{split}
		&\langle z(t_{I+1})|z(t_{I})\rangle=e^{-\frac{1}{2}\int |z_{I}|^2}e^{-\frac{1}{2}\int |z_{I+1}|^2} \left\langle 0\left|\exp\left\{\int \bar{z}_{I+1}\hat{a}_{I+1}\right\} \exp\left\{\int z_I\hat{a}^\dagger_I\right\}\right|0\right\rangle\\
  &=e^{-\frac{1}{2}\int |z_{I}|^2}e^{-\frac{1}{2}\int |z_{I+1}|^2} \left\langle 0\left|\exp\left\{\int \bar{z}_{I+1}(\hat{a}_{I}+\Delta t\left(i\omega_\bk\hat{a}_I\right))\right\} e^{-i\hat{H}\Delta t}\exp\left\{\int z_I\hat{a}^\dagger_I\right\}\right|0\right\rangle.
	\end{split}
\end{equation}}
where we used the time evolution of $e^{\int \bar{z} a}$.
Since the Hamiltonian with  interaction $V(t)\equiv V[a(\bk,t),a^{\dagger}(\bk,t);t]$, which is assumed to decay rapidly at future and past infinities, is
\begin{equation}
    \hat{H}(t)=\int d\tk ~\omega_\bk \hat{a}^\dagger(\bk,t)\hat{a}(\bk,t)+V(t),
\end{equation}
$\langle z(t_{I+1})|z(t_{I})\rangle$ becomes
\AM{\begin{equation}
    \begin{split}
  \langle z(t_{I+1})|z(t_{I})\rangle&=e^{-\frac{1}{2}\int |z_{I}|^2}e^{-\frac{1}{2}\int |z_{I+1}|^2}\exp{\int d\tk~\bar{z}_{I+1}z_I(1+i\omega_\bk\Delta t-i\omega_\bk\Delta t)}~e^{-iV(\bar{z}_{I+1},z_I)\Delta t}\\
  &= \exp\left\{-\int \dk \frac{\Delta t}{2}(\bar{z}_I\dot{z}_I-z_I\dot{\bar{z}}_I)-iV(\bar{z}_{I+1},z_I)\Delta t\right\}+\mathcal{O}((\Delta t)^2).
    \end{split}
\end{equation}}
Inserting this, we obtain the matrix element of the evolution operator in the coherent basis of the form 
\begin{equation}
	\begin{split}
		\langle f,t_f|i,t_i\rangle&=\prod_I\frac{\mathcal{D}z(t_I)\mathcal{D}\tilde{z}(t_I)}{ V}e^{-\int \dk \frac{\Delta t}{2}(\bar{z}_I\dot{z}_I-z_I\dot{\bar{z}}_I)-iV(t)\Delta t}\langle f|z(t_f)\rangle\langle z(t_i)|i\rangle\\\
		&\to \int \mathcal{D}z \mathcal{D}\bar{z }~e^{i\int dt\dk \frac{i}{2}(\bar{z}(\bk)\dot{z}(\bk)-z(\bk)\dot{\bar{z}}(\bk))-V(t)}\psi^*_f\psi_i.\label{path integral in coherent basis}
	\end{split}
\end{equation}
In order to converge the path integral, we implicitly multiply $\mathcal{H}$ by $1-i\epsilon ~(0<\epsilon\ll 1)$ ($\epsilon$-trick). 
Finally, we obtain \AM{a} generating function in the coherent path integral.
\begin{align}
	Z[J]:&= \int \mathcal{D}z \mathcal{D}\bar{z }~e^{i\int dt\dk~\mathcal{L}(\bk,t)+J(\bk,t)z(\bk,t)+\bar{J}(\bk,t)\bar{z}(\bk,t)},\\ \mathcal{L}(\bk,t)&=\frac{i}{2}(\bar{z}(\bk,t)\dot{z}(\bk,t)-z(\bk,t)\dot{\bar{z}}(\bk))-\mathcal{V}(\bk,t).
\end{align}
Using this generating function, we can write a scattering amplitude
\begin{align}
	\langle f|i\rangle=\left.\frac{\delta}{\delta J(\bk,\infty)}\dots\frac{\delta}{\delta\bar{J}(\bk,-\infty)}{Z[J]}/Z[0]\right|_{J=\bar{J}=0}.
\end{align}
 
\paragraph{Derivation of a scalar field's Lagrangian: From canonical  formalism}
We can also derive the above Lagrangian of a scalar field in another way: using canonical formalism. 
We start from the canonical Lagrangian 
\begin{equation}
	\mathcal{L} = \frac{1}{2} \partial_{\mu} \Psi \partial^{\mu} \Psi   - V(\Phi).
\end{equation} 
As before, we introduce the creation and annihilation operators by
\begin{equation}
	\begin{aligned}
		\Psi(X) &=\int \frac{d^3 \mathbf{k}}{2 \omega_{\mathbf{k}}}\left[a(\mathbf{k}, t) e^{i k \cdot X}+a^{\dagger}(\mathbf{k}, t) e^{-i k \cdot X}\right],\\
		\Pi(X)&=  \int \frac{d^3 \mathbf{k}}{2 \omega_{\mathbf{k}}} \, ( -i\omega_{\mathbf{k}}) \,\left[a(\mathbf{k}, t) e^{i k \cdot X} - a^{\dagger}(\mathbf{k}, t) e^{-i k \cdot X}\right].
	\end{aligned}\label{eq:canoniacalExpansion}
\end{equation}
In terms of these operators, the Hamiltonian is given by
\begin{equation}
	H = \int \frac{d^{3}x}{(2\pi)^{3}} \, \mathcal{H} = \frac{ 1 }{2} \int d\tilde{k} \, \omega_{\mathbf{k}} \left( a^{\dagger}(\mathbf{k}, t) a(\mathbf{k}, t) + a(\mathbf{k}, t) a ^{\dagger}(\mathbf{k}, t) \right) + V(a,a^{\dagger}).
\end{equation}
Using this Hamiltonian, the action is given by
\begin{equation}
	\begin{aligned}
		S_\text{Origi.}&=\int \frac{d t d^3 x}{(2 \pi)^3} \,  \left( \Pi \partial_{0} \Psi - \mathcal{H} \right)\\
		& =  \int_{I} dt \,  \int d\tilde{k} \, \frac{i}{2} \left( a^{\dagger}(\mathbf{k},t) \partial_{t}  a(\mathbf{k},t) -a(\mathbf{k},t) \partial_{t}  a^{\dagger}(\mathbf{k},t) \right) - \int_{I} dt \, V(a,a^{\dagger})+ S_\text{bdy},
	\end{aligned}
\end{equation}
where we explicitly introduced the boundary term $S_\text{bdy}$ by 
\begin{equation}
\begin{aligned}
		S_\text{bdy}
		&= - \frac{i}{4} \int d\tilde{k} \, \int_{t\in \partial I} \left[a(\mathbf{k},t) a(-\mathbf{k},t) e^{-2i\omega_{\mathbf{k}} t} -a^{\dagger}(\mathbf{k},t) a^{\dagger}(-\mathbf{k},t) e^{2i\omega_{\mathbf{k}} t}  \right].
\end{aligned}
\end{equation}


We note that the existence of the boundary term $S_\text{bdy}$ plays an important role when we consider path integral with the non-trivial initial and final states. To see the role of the boundary term more explicitly, let us consider the path integral in the canonical variables. In this case, the initial and final wave functionals corresponding to the vacuum state in a free theory are given by
\begin{equation}
	\begin{aligned}
		\psi_{\alpha} &= \exp\left[  -\frac{1}{4} \int \frac{d\tilde{k}}{(2\pi)^{6}}  (2\omega_{\mathbf{k}})^{2} \, \tilde{\Psi}(t_{\alpha},\mathbf{k}) \tilde{\Psi}(t_{\alpha},-\mathbf{k})   \right]\\
				& =\exp\left[  -\frac{1}{4} \int d\tilde{k} \left\{ \left( a(\mathbf{k}, t_{\alpha}) a^{\dagger}(-\mathbf{k}, t_{\alpha}) + a(\mathbf{k}, t_{\alpha}) a(-\mathbf{k}, t_{\alpha})e^{ - 2i \omega_{\mathbf{k}} t_{\alpha}} \right) +(\text{h.c.})  \right\}  \right] \qquad (\alpha = \text{in, out}),
	\end{aligned}
\end{equation}
where $t_{f,i}$ correspond to the boundary times $\partial I=\{ t_f,t_i \}$. Combining this wave functionals $\psi_f^{*} \, \psi_i$ and the boundary term $iS_\text{bdy}$, we get the modified boundary term by the boundary states in the path integral (see e.g., \cite{Marolf:2004fy}),
\AM{\begin{equation}
	\begin{aligned}
		i \widetilde{S}_\text{bdy,Path Integral}&= iS_\text{bdy} + \log \psi_i + \log \psi_f^{*}\\
		&=-  \frac{1}{4} \sum_{\alpha=\{ f,i \} } \int d\tilde{k}  \left\{  a(\mathbf{k}, t_{\alpha}) a^{\dagger}(-\mathbf{k}, t_{\alpha}) +(\text{h.c.})  \right\} \\
		& \qquad \quad - \frac{1}{2} \int d\tilde{k} \left[ a^{\dagger}(\mathbf{k},t_f) a^{\dagger}(-\mathbf{k},t_f) e^{2i\omega_{\mathbf{k}} t_f} +a(\mathbf{k}, t_i) a(-\mathbf{k}, t_i) e^{-2i\omega_{\mathbf{k}}  t_i}   \right].
	\end{aligned}
\end{equation}}
The above boundary terms in the second line vanish due to the boundary conditions for $a(\mathbf{k},t_{i})$, $a^{\dagger}(\mathbf{k},t_{f})$ in the vacuum state \footnote{See the boundary conditions \eqref{eq:bdycondVacum}.}. 



\paragraph{two-point scattering}

With \AM{Fourier} transforming of time
\AM{\begin{align}
    z(\bk,t)&=\frac{1}{2\pi}\int dl~e^{-ilt}\tilde{z}(l,\bk),\\
    \bar{z}(\bk,t)&=\frac{1}{2\pi}\int dm~e^{imt}\tilde{\bar{z}}(l,\bk),
\end{align}}
the action with source terms becomes
\begin{equation}
    \begin{split}
        iS[J,\bar{J}]&=\int\!\dk dt~\left[\frac{-1}{2}\left(\bar{z}(\bk,t)\dot{z}(\bk,t)-z(\bk,t)\dot{\bar{z}}(\bk,t)\right)\right.\\
        &\left.\qquad +2\omega_\bk\bar{J}(\bk,t)\bar{z}(\bk,t)+2\omega_\bk{J}(\bk,t){z}(\bk,t)\right]\\
        &=\int\!\dk \frac{dl}{2\pi}~\left[\tilde{z}(\bk,l)\tilde{\bar{z}}(\bk,l)\left(il-\epsilon\right)+2\omega_\bk\left(\tilde{z}(\bk,l)\tilde{J}(\bk,l)+\tilde{\bar{z}}(\bk,l)\tilde{\bar{J}}(\bk,l)\right)\right].
    \end{split}
\end{equation}
Integrating out by $\tilde{z}$ and $\tilde{\bar{z}}$ the partition function is
\AM{\begin{equation}
    \begin{split}
        Z[J]&\propto\exp{\int d^3\bk\frac{dl}{2\pi i}~\frac{(2\omega_\bk) \tilde{J}(\bk,l)\tilde{\bar{J}}(\bk,l)}{l+i\epsilon}},
    \end{split}
\end{equation}}
\AM{where we changed variables of the integral measure from $\md{z}{\bar{z}}$ to $\md{\tilde{z}}{\tilde{\bar{z}}}$.
Because} $a_\tout$ corresponds to $z_\tout=\frac{1}{2\pi}\int dl~e^{-ilT_\tout}\tilde{z}_\tout(l,\bk)$ and $a^\dagger_\tin$ corresponds to $\bar{z}_\tin=\frac{1}{2\pi}\int dm~e^{imT_\tin}\tilde{\bar{z}}_\tin(l,\bk)$ in the path integral, scattering amplitude can be  written as 
\AM{\begin{equation}
    \begin{split}
        \langle a_\tout(\bk_\tout) a_\tin^\dagger(\bk_\tin)\rangle
        &= \left.\lim_{T_\tin\to-\infty}\!\lim_{T_\tout\to \infty}\int\! {dl}{dm}~e^{i(mT_\tin-lT_\tout)}\frac{\delta^2}{\delta \bar{J}(l,\bk_\tout) \delta J(m,\bk_\tin)}\right|_{\tilde{J}=\tilde{\bar{J}}=0}\!\log[Z[J]]\\
        &=\lim_{T_\tin\to-\infty}\!\lim_{T_\tout\to \infty}\int\! {dl}~e^{il(T_\tin-T_\tout)}\frac{(2\omega_\bk)}{(l+i\epsilon)(2\pi i)}\delta^3(\bk_\tin-\bk_\tout)\\
        &=\lim_{T_\tin\to-\infty}\!\lim_{T_\tout\to \infty}(2\omega_\bk)\delta^3(\bk_\tin-\bk_\tout)\Theta(T_\tout-T_\tin),
    \end{split}
\end{equation}}
where we used the $\epsilon$-trick $\mathcal{H}\to\mathcal{H}(1-i\epsilon)$.
\subsection{Path integral in Celestial holography for free theory}
Massless celestial holography claims that the Mellin transform of the bulk scattering amplitude of $n$-particles is equivalent to the correlation function of the $n$-point celestial operators. We will formulate the path integral of this celestial holography.
\par 
Defining
\AM{\begin{align}
    \alpha_\Delta(t,w,\bar{w})&:=\int d\omega~\omega^{\Delta-1}z(t,\bk(\omega,w,\bar{w}))\quad \to \quad z(\bk,t)=\int\!\frac{d\Delta}{2\pi i}~\omega^{-\Delta}\alpha_\Delta(t,w,\bar{w}),\\
    \bar{\alpha}_\Delta(t,w,\bar{w})&:=\int d\omega~\omega^{\Delta-1}\bar{z}(t,\bk(\omega,w,\bar{w}))\quad \to \quad \bar{z}(\bk,t)=\int\!\frac{d\Delta}{2\pi i}~\omega^{-\Delta}\bar{\alpha}_\Delta(t,w,\bar{w}),
\end{align}}
we can formulate massless Celestial CFT path integral by inserting the complete-basis of asymptotic boundary states into the generating function as follows
\color{blue}
\color{black}
\begin{equation}
    \begin{split}
         Z[J_f,J_i]=& \int \md{\bar{\alpha}_f}{\alpha_f}\md{\bar{\alpha}_i}{\alpha_i}~\langle 0|\alpha_f\rangle\\
    &\times\left\langle \alpha_f\left|\prod_{t}\int\md{z(t)}{\bar{z}(t)}|z(t)\rangle\langle \bar z(t)|
    \right|\alpha_i\right\rangle \langle\alpha_i|0\rangle\\
    =&\int \md{\bar{\alpha}_f}{\alpha_f}\md{\bar{\alpha}_i}{\alpha_i}~\langle 0|\alpha_f\rangle\\
    &\times\left\langle \alpha_f\left|\prod_{t}\int\md{\alpha(t)}{\bar{\alpha}(t)}|\alpha(t)\rangle\langle \bar\alpha(t)|
    \right|\alpha_i\right\rangle \langle \alpha_i|0\rangle\\
    =& \int \md{^2{\alpha}_i}{^2\alpha_f}~\langle 0|\alpha_f\rangle\langle \bar\alpha_i|0\rangle\int\md{\bar{\alpha}}{\alpha}\left\langle \bar\alpha_f|\alpha(t_f)\rangle\langle\bar\alpha(t_i)|{\alpha}_i\right\rangle .\label{eq:ccft path integral}
    \end{split}
\end{equation}
Meanwhile, we can take the complete basis of asymptotic states as
\AM{\begin{equation}
   \begin{split}
        |\alpha_\Delta\rangle=|z[\alpha]\rangle&=e^{-\frac{1}{2}\int d\tilde{k}|z(\bk,t)|^2} e^{\int d\tk~z(\bk,t)\hat{a}^\dagger(\bk,t)}|0\rangle \\
        &=e^{-\frac{1}{2}\int d^2w\frac{d\Delta}{2\pi i} \alpha_\Delta\bar{\alpha}_{\Delta^*}}e^{\int d^2w\frac{d\Delta}{2\pi i}~ \alpha_{\Delta^*}(t) \hat{O}^\dagger_\Delta(t)}|0\rangle,
   \end{split}
\end{equation}}
by defining
\AM{\begin{align}
    \hat{O}^{\dagger}_{\Delta}(t,w,\bar{w})&=\int d\omega~\omega^{\Delta-1}\ha^\dagger(t,\bk(\omega,w,\bar{w})),\\
    \hat{O}_\Delta(t,w,\bar{w})&=\int d\omega~\omega^{\Delta-1}\ha(t,\bk(\omega,w,\bar{w})),
\end{align}}
which satisfy the commutation relation for free theory
\begin{align}
    [\hat{O}_\Delta(t,w,\bar{w}),\hat{O}^\dagger_{\Delta'}(t,w',\bar{w}')]=2\pi i\delta(\Delta-\Delta'^*)\delta^2(w-w').
\end{align}
\AMM{Here, we do not treat the operators $\hat{O}^{\dagger}_{\Delta}(t,w,\bar{w})$ and $\hat{O}_\Delta(t,w,\bar{w})$ as Celestial CFT operators, but rather as bulk operators\footnote{\AMM{We do not know such objects can be described by degrees of freedom of Celestial CFT. It would be interesting to find such a description.}}. On the other hand, we treat the operators $\hat{O}^{\Delta}_{i}(w,\bar{w})$ and $\hat{O}^\Delta_{f}(w,\bar{w})$, which are used implicitly to define the states $\langle \bar{\alpha}_{i}|$ and $| \alpha_{f} \rangle$, as Celestial CFT operators.}, where they are defined at $t_{i} \leq t \leq t_{f}$.

\AM{Then, we obtain}
\begin{align}
    Z=& \int \md{^2{\alpha}_i}{^2\alpha_f}~e^{\int-|\alpha_i|^2-|\alpha_f|^2}\\
    &\quad \times \int\mathcal{D}^2{\alpha} ~e^{iS[\alpha,\bar{\alpha}]+\int -\frac{1}{2}|\alpha(t_f)|^2-\frac{1}{2}|\alpha(t_i)|^2+\bar{\alpha}_f\alpha(t_f)+\bar{\alpha}(t_i)\alpha_i+\bar{J}_i\bar{\alpha}_i+{J}_f\alpha_f}\notag.
\end{align}
\AM{We} conclude that the partition function is generally 
\AM{\begin{equation}
    \begin{split}
    Z=\int\mathcal{D}^2\alpha(t_f)\mathcal{D}^2\alpha(t_i)~e^{iS_{bdy,i}[\alpha_i,\bar{\alpha}_i]+iS_{bdy,f}[\alpha_f,\bar{\alpha}_f]+\int -\frac{1}{2}|\alpha(t_f)|^2-\frac{1}{2}|\alpha(t_i)|^2+\bar\alpha(t_i)\bar{J}_i+{\alpha}(t_f)J_f}.
    \end{split}
\end{equation}}
\paragraph{Two-point correlator for a free scalar theory}
Let us now compute the two-point Celestial CFT correlator for \AM{a free scalar field} in the dual bulk. The action is
\begin{equation}
    \begin{split}
         S&=\int\!\dk dt~\frac{i}{2}\left(\bar{z}(\bk,t)\dot{z}(\bk,t)-z(\bk,t)\dot{\bar{z}}(\bk,t)\right)\\
    &=\int\!\dk dt~\frac{i}{2}\left(\int\!\frac{d\Delta}{2\pi i}\int\!\frac{d\Delta'}{2\pi i}~\omega^{-\Delta}\bar{\alpha}_\Delta(t,w,\bar{w})\omega^{-{\Delta'}}\dot{\alpha}_{\Delta'}(t,w,\bar{w})\right.\\
    &\hspace{3.4cm} \left.-\int\!\frac{d\Delta}{2\pi i}\int\!\frac{d\Delta'}{2\pi i}~\omega^{-\Delta}\dot{\bar{\alpha}}_\Delta(t,w,\bar{w})\omega^{-{\Delta'}}{\alpha}_{\Delta'}(t,w,\bar{w})\right)\\
    &=\frac{i}{2}\int \! dtd^2w\frac{d\Delta}{2\pi i}\left(\bar{\alpha}_\Delta(t,w,\bar{w})\dot{\alpha}_{\Delta^*}(t,w,\bar{w})-\dot{\bar{\alpha}}_{\Delta^*}(t,w,\bar{w}){\alpha}_{\Delta}(t,w,\bar{w})\right).
    \label{action:coherent}
    \end{split}
\end{equation}
Substituting the solutions of equations of motion
\AM{\begin{align}
    \alpha_\Delta(t)=\alpha_\Delta(0),\quad \bar{\alpha}_\Delta(t)=\bar{\alpha}_\Delta(0),
\end{align} }
into the \AM{action }\eqref{action:coherent}, we can evaluate the path \AM{integral }\eqref{eq:ccft path integral} as
\begin{equation}
\begin{split}
    Z[\bar{J}_i,J_f]&=\int\md{^2{\alpha}_i}{^2\alpha_f}~\langle \bar{\alpha}_i|0\rangle\langle 0|\alpha_f\rangle \int \md{\bar{\alpha}(0)}{{\alpha}(0)}\left\langle \bar{\alpha}_f|\alpha(t_f)\rangle\langle\bar{\alpha}(t_i)|{\alpha}_i\right\rangle e^0e^{J_f\alpha_f+\bar{J}_i\bar{\alpha}_i}\\
    &=\int\md{^2{\alpha}_i}{^2\alpha_f}~e^{\int~-{|{\alpha}_i|^2}-{|{\alpha}_f|^2}} ~\\
    &\quad \times \int \md{\bar{\alpha}(0)}{{\alpha}(0)}~e^{\int \!d^2w\frac{d\Delta}{2\pi i }~\bar{\alpha}_f\alpha(0)+\bar{\alpha}(0)\alpha_i-|\alpha(0)|^2+J_f\alpha_f+\bar{J}_i\bar{\alpha}_i}\\
    &\propto\int\md{^2{\alpha}_i}{^2\alpha_f}~e^{\int~{\alpha}_i\bar\alpha_f-|\alpha_f|^2-|\alpha_i|^2+J_f\alpha_f+\bar{J}_i\bar{\alpha}_i}.
\end{split}
\end{equation}
Substituting $\bar{\alpha}_f=J_f$ and $\alpha_i=\bar{J}_i$, 
\begin{equation}
   Z[\bar{J}_i,J_f] \propto~\exp\left[{\int\frac{d\Delta}{2\pi i}d^2w~\bar{J}^\Delta_i(w,\bar{w}){J}^{\Delta^*}_f(w,\bar{w})}\right].
\end{equation}
By taking functional derivatives with respect to $\bar{J}_i,~J_f$, we obtain
\begin{align}
    \langle \hat{O}^\Delta_f (w,\bar{w})\hat{O}^{\Delta'}_i(w',\bar{w}')\rangle=2\pi i\delta(\Delta-\Delta'^*)\delta^2(w-w').
\end{align}

\section{C-function from bulk partition function}\label{appendix:cfunction}

In this appendix, we derive a c-function of Wedge CFT  in the same way as \AM{AdS/CFT }\cite{Gubser:2002vv,Hartman:2006dy}.


A c-function is defined in \AM{qunatum field theory (QFT)} and it is determined from the QFT partition. In CFT, the c-function coincides with a central charge. 
In AdS/CFT, since the CFT partition function is equal to the corresponding bulk partition function through the GKPW dictionary \cite{Hartman:2006dy}, one can evaluate the c-function from bulk side. 
 The same logic can be used for wedge-holography case. We can evaluate a central charge of Wedge CFT, assuming GKPW-like dictionary \eqref{GKPW}.

First, the c-function can be determined from the partition function $W_{\gamma}$ as follows, e.g., \cite{Hartman:2006dy},
\begin{equation}
c_{\gamma}^{\text{matter}} = \frac{1}{2}\tilde{R}\frac{\partial}{\partial \tilde{R}} W_{\gamma}\left[\tilde{R}\right],
\end{equation}
where $\tilde{R}$ is a radius of the sphere on which the boundary theory is defined. Here, we focus only on corrections coming from matters related to the scalar field discussed in subsection \ref{subsec:changingbdyCondi}\footnote{In general, the c-function can be defined by adding contributions from the Einstein-Hilbert action. However, such contributions are not dominant contributions to the change of the c-function under changes of boundary conditions (or double-trace deformations), and dominant ones come from matter fields. See \AM{e.g.,} \cite{Gubser:2002vv}. Thus, we also focus on the matter contributions to the c-function here.}. 
For the current setup, the partition function is given by
\begin{equation}
	W_{\gamma}\left[\tilde{R}\right]= -\frac{1}{2} \int d\Delta \sum_{l=0}^{\infty} (2l+1) \log \left[ (\gamma-1)\tilde{R}^{\epsilon}+(\gamma -\Delta)\tilde{R}^{-2(\Delta -1)} \cdot \frac{\Gamma[1-\Delta]}{\Gamma[\Delta-1]} \cdot \frac{\Gamma[l+\Delta]}{\Gamma[l+2-\Delta]}   \right],
\end{equation}
where $\tilde{R}$ is related to $R$ through $\tilde{R}=e^{R}$.  This expression can be derived by integrating our $O_{\Delta}$ correctly in the on-shell action \eqref{eq:MasslessActionOnAdSPatch}.

Generally, it is difficult to evaluate the above expression analytically for arbitrary $\gamma$.
Here, we focus on $\gamma=1$ which is a very important case as seen in equation \eqref{eq:twogamma1}. In this case, $\tilde{R}$-dependent terms of the partition function becomes
\begin{equation}
	W_{\gamma=1}\left[\tilde{R}\right]= -\frac{1}{2} \int d\Delta \sum_{l=0}^{\infty} (2l+1) \log \left[\tilde{R}^{-2(\Delta -1)} \cdot \frac{\Gamma[l+\Delta]}{\Gamma[l+2-\Delta]}   \right] + (\tilde{R}\text{-independent terms}),
\end{equation}
where we separated $\tilde{R}$-independent terms which are not essential to obtain the central charge. 
In evaluating the above partition function, we need to consider the two divergent sums
\begin{equation}
	\sum_{l=0}^{\infty} (2l+1), \qquad \sum_{l=0}^{\infty} (2l+1) \log\left[ \frac{\Gamma[l+\Delta]}{\Gamma[l+2-\Delta]} \right].
\end{equation}
For AdS/CFT cases, evaluations of these sums are discussed in \cite{Gubser:2002vv}, and we can use the same results. 
Then, the $\log\tilde{R}$-divergent terms for the divergent sum can be given by
\begin{equation}
	\begin{aligned}
		&\sum_{l=0}^{\infty} (2l+1) \log \left[\tilde{R}^{-2(\Delta -1)} \cdot \frac{\Gamma[l+\Delta]}{\Gamma[l+2-\Delta]}   \right]\\
		&= -2(\Delta-1)\log \tilde{R} \sum_{l=0}^{\infty}(2l+1) + \sum_{l=0}^{\infty} (2l+1) \log \left[\frac{\Gamma[l+\Delta]}{\Gamma[l+2-\Delta]}   \right]\\
		& \to  -2(\Delta-1)\log \tilde{R} \cdot \frac{1}{3} +  \left( \frac{2}{3} (\Delta-1) \log\tilde{R}  - \frac{2(\Delta-1)^{3}}{3} \log\tilde{R}\right)\\
		& =  - \frac{2(\Delta-1)^{3}}{3} \log\tilde{R}.
	\end{aligned}
\end{equation}
Then, the $\log\tilde{R}$-divergent part of the partition function is given by
\begin{equation}
	W_{\gamma=1}\left[\tilde{R}\right] \to \log\tilde{R} \int d\Delta \frac{(\Delta-1)^{3}}{3},
\end{equation}
and this result implies the c-function for $\gamma=1$,
\begin{equation}
	c^\text{matter}_{\gamma=1}=\int d\Delta \frac{(\Delta-1)^{3}}{6}.\label{eq:c-function}
\end{equation}
Here, we might be able to define c-functions for each $\Delta$
\begin{equation}
    c^\text{matter}_{\Delta,\gamma=1}=\frac{(\Delta-1)^{3}}{6},
\end{equation}
and each of these c-functions is associated to \AM{the} result obtained in \cite{Gubser:2002vv}.
The above results can be also expressed as
\begin{equation}
	c^\text{matter}_{\gamma=1}= \frac{1}{6}\int_{\mathbb{R}} d\nu\, \nu^{3}=0,
\end{equation}
The integrated c-functions are vanishing due to oddness of the integrand, $\nu^{3}$, although for each $\Delta=1+i\nu$, c-functions are non-zero.
We note that each c-functions for $\Delta$ is pure imaginary.

Here, we also note that we focus only on matter contributions to the c-function. The dominant contribution to the c-function which comes from the Einstein-Hilbert action is expected to pure imaginary divergent central charge, as observed in earlier \AM{literature }\cite{Pasterski:2022lsl,Ogawa:2022fhy} which used different approaches. 

%

%

%
%

\section{Bulk two point functions in Harmonic basis}\label{appendix:harmonic basis}
In this appendix, we explain another method of solving the mixed boundary condition \eqref{eq:bdy cond. massless} by using a spherical harmonic basis in massless and massive scalars. To use this method, it is convenient to rewrite the scalar field using the spherical harmonic function explicitly. Such an expression for massive cases is discussed in \cite{Ogawa:2022fhy}, and using the expression, we can write the scalar field as
\begin{equation}
    \Psi_{lm}(\eta,\rho)=\int d^2\Omega_0 f_{\Delta}(\eta) \frac{\mathcal{N}}{ \sinh\rho }  Y^*_{lm}(\Omega_0)Q^{\Delta-1}_l(\coth\rho)\phi_\Delta(\Omega_0),
\end{equation}
where $f_{\Delta}(\eta)$ is given by\footnote{This function is related to $\eta$-dependent factors in conformal primary wavefunction. See \ref{subsec:Celestialholography}. }
\begin{equation}
f_{\Delta}(\eta)= \begin{cases}
        \eta^{1-\Delta} & \text{for massless case}\\
         H_{\Delta}^{(\pm)}(m\eta) & \text{for massive case}\\
    \end{cases}
\end{equation}
and $\mathcal{N}$ is given by $ \mathcal{N}= e^{i(1-\Delta)\pi}$.

The above expression is obtained by considering the inner product between the scalar field $\Psi(\eta,\rho,\Omega)$ and the spherical harmonic function $Y_{lm}$,
\begin{equation}
     \Psi_{lm}(\eta,\rho) = \int d^2\Omega_0 \, Y_{l,m}^{*}(\Omega_{0})  \Psi(\eta,\rho,\Omega_{0}).
\end{equation}

Let us solve the boundary value problem using the above expression.
In the spherical harmonic basis, the mixed boundary condition \eqref{eq:bdy cond. massless} and the scalar field expansion at $\rho\gg 1$ are written as
\AM{\begin{align}
	\gamma\Psi^\Delta_{lm}+\partial_\rho\Psi^\Delta_{lm}
|_{\rho\to\infty}&=\tilde{\phi}^\Delta_{b,lm}(\eta),\\
	(\Psi|_{\rho\to\infty})^\Delta_{lm}(\eta)&=\int d^2\Omega_0~f_\Delta(\eta)\frac{\mathcal{N}}{\sinh\rho}~Y^*_{lm}(\Omega_0)Q^{\Delta-1}_l(\coth\rho)\chi_\Delta(\Omega_0),
\end{align}}
respectively.
Combining them, we can rewrite $\tilde{\phi}$ by boundary field $\tilde{\phi}_b(\Omega)$ as follows
\begin{align}
	\tilde{\phi}^\Delta_{b,lm}(\eta)=&\partial_\rho\left[\frac{\mathcal{N}}{\sinh\rho}\int d^2\Omega_0{Y^*_{lm}(\Omega_0)}{f_\Delta(\eta)Q_{l}^{\Delta-1}(\coth\rho)}\chi(\Omega_0)\right]\notag\\
	&\qquad +\frac{\gamma\mathcal{N}}{\sinh\rho}\int d^2\Omega_0~{Y^*_{lm}(\Omega_0)}{f_\Delta(\eta)Q_{l}^{\Delta-1}(\coth\rho)}\chi_\Delta(\Omega_0)\notag\\
	\overset{\rho\to \infty}{\simeq} & {\mathcal{N}}\int d^2\Omega_0{e^{i(\Delta-2)\pi}}Y^*_{lm}(\Omega_0)f_\Delta(\eta)\notag\\
	&\qquad \times\left[(\gamma+\Delta-2)\Gamma[\Delta-1]e^{(\Delta-2)\rho}+(\gamma-\Delta)\frac{\Gamma[1-\Delta]\Gamma[\Delta+l]}{\Gamma[l+2-\Delta]}e^{(-\Delta)\rho}\right]\chi_\Delta(\Omega_0),\\
	\rightarrow \chi_\Delta(\Omega)=&\sum_{l,m}Y_{lm}(\Omega)\frac{e^{i(1-\Delta)\pi}}{{\mathcal{N}}f_\Delta(\eta)\left[(\Delta-2+\gamma)\Gamma[\Delta-1]e^{(\Delta-2)\rho}+(\gamma-\Delta)\frac{\Gamma[1-\Delta]\Gamma[\Delta+l]}{\Gamma[l+2-\Delta]}e^{-\Delta\rho}\right]}\tilde{\phi}^\Delta_{b,lm}.
\end{align}
Thus, the action becomes
\begin{align}
	S_{EOM}=&\int  d\eta d^2\Omega\frac{d\Delta}{2\pi i}\frac{d\Delta'}{2\pi i}~\sqrt{-g}\Psi(\eta,\rho,\Omega)\phi_b(\Omega;\eta,\rho)|_{\rho=\rho_\infty}\\
	=&\int \frac{f_\Delta(\eta)f_{\Delta'}(\eta)}{\eta}e^{2\rho}\sum_{l,m}\frac{\Gamma[\Delta'-1]e^{(\Delta'-2)\rho}+\frac{\Gamma[1-\Delta']\Gamma[\Delta'+l]}{\Gamma[l+2-\Delta']}e^{(-\Delta')\rho}}{(\Delta-2+\gamma)\Gamma[\Delta-1]e^{(\Delta-2)\rho}+(\gamma-\Delta)\frac{\Gamma[1-\Delta]\Gamma[\Delta+l]}{\Gamma[l+2-\Delta]}e^{(-\Delta)\rho}}\tilde{\phi}^\Delta_{b,lm}\tilde{\phi}^\Delta_{b,lm}.
\end{align}
For a massless scalar, since we have $\int \frac{f_\Delta(\eta)f_{\Delta'}(\eta)}{\eta}\sim \delta(\Delta+\Delta'-2)$ picks up $\Delta'=2-\Delta$ \AM{mode,} two-point function is
\AM{\begin{align}
	G^{lm,no}&=\delta_{l,n}\delta_{m,o}\frac{\Gamma[1-\Delta]e^{(-\Delta)\rho}+\frac{\Gamma[\Delta-1]\Gamma[2-\Delta+l]}{\Gamma[l+\Delta]}e^{(\Delta-2)\rho}}{(\Delta-2+\gamma)\Gamma[\Delta-1]e^{(\Delta-2)\rho}+(\gamma-\Delta)\frac{\Gamma[1-\Delta]\Gamma[\Delta+l]}{\Gamma[l+2-\Delta]}e^{(-\Delta)\rho}}\notag\\
	&=\delta_{l,n}\delta_{m,o}\frac{G_0^l}{1+\gamma\frac{\Gamma [\Delta+l]}{[l+2-\Delta]}G_0^l},
\end{align}}
where 
\begin{equation}
	G_0^l=\frac{\Gamma[2-\Delta+l]}{\Gamma[l+\Delta]}\frac{\Gamma[\Delta-1]e^{(\Delta-2)\rho}+\frac{\Gamma[1-\Delta]\Gamma[\Delta+l]}{\Gamma[l+2-\Delta]}e^{(-\Delta)\rho}}{(\Delta-2)\Gamma[\Delta-1]e^{(\Delta-2)\rho}+(-\Delta)\frac{\Gamma[1-\Delta]\Gamma[\Delta+l]}{\Gamma[l+2-\Delta]}e^{(-\Delta)\rho}}.
\end{equation}
For a massive scalar, restricting to $\nu>0$, the two-point function is
\AM{\begin{align}
	G^{lm,no}&=\frac{\Gamma[\Delta-1]e^{(\Delta-2)\rho}+\frac{\Gamma[1-\Delta]\Gamma[\Delta+l]}{\Gamma[l+2-\Delta]}e^{(-\Delta)\rho}}{(\Delta-2+\gamma)\Gamma[\Delta-1]e^{(\Delta-2)\rho}+(\gamma-\Delta)\frac{\Gamma[1-\Delta]\Gamma[\Delta+l]}{\Gamma[l+2-\Delta]}e^{(-\Delta)\rho}}\notag\\
	&=\delta_{l,n}\delta_{m,o}\frac{G_0^l}{1+\gamma G_0^l},
\end{align}}
where
\begin{equation}
	G^l_0=\frac{\Gamma[\Delta-1]e^{(\Delta-2)\rho}+\frac{\Gamma[1-\Delta]\Gamma[\Delta+l]}{\Gamma[l+2-\Delta]}e^{(-\Delta)\rho}}{(\Delta-2)\Gamma[\Delta-1]e^{(\Delta-2)\rho}+(-\Delta)\frac{\Gamma[1-\Delta]\Gamma[\Delta+l]}{\Gamma[l+2-\Delta]}e^{(-\Delta)\rho}}.
\end{equation}


\section{Holography through the EOW brane}\label{appendix:holograEOW}
In this appendix, we explain an example of a holography reproducing an expected celestial CFT result by imposing certain exotic boundary conditions. Under the conditions, we will evaluate two-point functions using the techniques explained in the main body of this paper, and see that it indeed reproduces an expected celestial CFT two-point function.

In the second half of the main body of this paper, we have considered holography at $\rho\gg 1$, and correspondingly we have imposed boundary conditions at $\rho\gg 1$. Here, we impose boundary conditions on other boundary regions: EOW-branes.
Even the way of imposing boundary condition is different from the conventional one, we can derive the GKPW-like relation and evaluate two-point functions. Let us first give the GKPW-like relation.
\paragraph{Proof of GKPW}
The action with the mixed boundary condition on $\eta=\eta_i~(i=1,2)>0$ for region $\mathcal{A}^+$ and $r=R\gg 1$ for region D can be written as follows
\AM{\begin{equation}
	\begin{split}
		S&=S_{\mathcal{A}^+}+S_{\mathcal{D}}\\
  =&\int d^2\Omega d\eta d\rho~\eta^3\sinh^2\rho\frac{1}{2}\partial\Psi\partial\Psi\\
  &\quad-\int_{\eta=\eta_2} d\rho d^2\Omega~\eta^2\sinh^2\rho[J_2\Psi+\frac{f}{2}\Psi^2]+\int_{\eta=\eta_1} d\rho d^2\Omega~\eta^2\sinh^2\rho[J_1\Psi+\frac{f}{2}\Psi^2]\\
  &\qquad +\int d^2\Omega dr dt~r^3\sinh^2t\frac{1}{2}\partial\Psi\partial\Psi+\int_{r=R} dt d^2\Omega~r^2\sinh^2t[J'\Psi+\frac{f}{2}\Psi^2].
	\end{split}
\end{equation}}
Varying of this action with respect to the field $\Psi$, we obtain an equation of motion for the field and the boundary conditions
\begin{align}
    \left.\left[\eta_i\partial_\eta+{f_i}\right]\Psi\right|_{\eta=\eta_i}=-J_i,\label{eq:bdyCondiEta}\\ 
    \left.\left[r\partial_r+{f}\right]\Psi\right|_{r=R}=-J'.\label{eq:bdyCondiR}
\end{align}

Because the field $\Psi$ satisfies the equation of motion, as in the discussion of sections \ref{sec:operatorcorresp.} and \ref{subsec:CCFTvsWCFT}, we can rewrite $\Psi$ in terms of Celestial operators $J_i$ and $J'$. Expanding $J$ and $J'$ as 
\begin{align}
    J_i=\int d^{2}z\int \frac{d\Delta}{2\pi i}~\eta_{i,\mp}^{-\Delta}\phi^\pm_\Delta\hat{j}^i_\Delta,\\
     J'=\int d^{2}z\int \frac{d\Delta}{2\pi i}~r^{-\Delta}\phi^\pm_\Delta\hat{j}'_\Delta,
\end{align}
 the scalar field satisfying the boundary conditions \eqref{eq:bdyCondiEta} and \eqref{eq:bdyCondiR} can be constructed as follows;
 \begin{align}
	\Psi|_{\eta=\eta_i}&=\int d^{2}z \int \frac{d\Delta}{2\pi i}~\frac{1}{\Delta-f_i}\eta_i^{-\Delta}\phi_\Delta (\rho,w,\bar{w};z,\bar{z}) \hat{j}^i_\Delta (z,\bar{z}),\\
	 \qquad \Psi|_{r=R}&=\int d^{2}z\int \frac{d\Delta}{2\pi i}~\frac{1}{\Delta-f} r^{-\Delta}\phi_\Delta (t,w,\bar{w};z,\bar{z}) \hat{j}'^i_\Delta (z,\bar{z}).
\end{align}


Now, using the above relations, we can prove the GKPW relation as discussed in \ref{subsec:changingbdyCondi}. Varying the action, we have
\begin{equation}
	\begin{split}
		\delta_{j_\Delta} S=&\int_{\eta=\eta_1} \hspace{-15pt}d\rho d^2\Omega ~\eta^2\sinh^2\!\rho~\delta J~\Psi-(1\leftrightarrow 2)+\int_{r_\infty}dt d^2\Omega ~r^2\sinh^2\!t~\delta J~\Psi\\
		=&\int_{\eta=\eta_1} \hspace{-15pt} d\rho d^2\Omega ~\eta^3\sinh^2\rho\int \frac{d\Delta}{2\pi i} \frac{d\Delta'}{2\pi i} ~\eta^{-\Delta-\Delta'-1}\int d^2z d^2 z' \\
  &\qquad\times\sum_{a=\pm,b=\pm}\phi^a_{\Delta'}(\rho,w,\bar{w};z',\bar{z}')\phi^b_\Delta(\rho,w,\bar{w};z,\bar{z})\delta j_{\Delta'}^a(z',\bar{z}')\hat{O}^b_\Delta(z,\bar{z})-(1\leftrightarrow 2)\\
  &+\int_{r_\infty} \hspace{-5pt} dt d^2\Omega ~r^3\sinh^2t\int \frac{d\Delta}{2\pi i} \frac{d\Delta'}{2\pi i} ~r^{-\Delta-\Delta'-1}\int d^2z d^2 z' \\
  &\qquad\times\sum_{a=\pm,b=\pm}\phi^a_{\Delta'}(\rho,w,\bar{w};z',\bar{z}')\phi^b_\Delta(\rho,w,\bar{w};z,\bar{z})\delta j_{\Delta'}^a(z',\bar{z}')\hat{O}^b_\Delta(z,\bar{z})\\
		=&\int \frac{d\Delta}{2\pi i} \frac{d\Delta'}{2\pi i}\frac{1}{\Delta'-\Delta}\int d^2z ~\hat{O}^+_\Delta(z,\bar{z})\delta j^-_{\Delta'}(z,\bar{z})\times \frac{8\pi^4 i}{(2\pi)^3}\delta(\Delta+\Delta'-2)+(+\leftrightarrow -)
	\end{split}
\end{equation}
where, in moving from the second equality to the third one, we used 
\AM{\begin{equation}
    \eta^{1-\Delta-\Delta'}=\frac{[\eta^{-\Delta}\partial_\eta{\eta^{-\Delta'}}-\eta^{-\Delta'}\partial_\eta{\eta^{-\Delta}}]}{\Delta'-\Delta},
\end{equation}}
and the orthonormality of conformal primary wavefunctions in the Klein-Gordon inner product. Thus, the GKPW relation is
\begin{equation}
	\begin{split}
		\langle O^\pm_\Delta(z,\bar{z})\dots \rangle \propto \left.\frac{\delta }{\delta j^\mp_{2-\Delta}}\dots \right|_{j=0}S.
	\end{split}
\end{equation}
\qed 
%
%
%
\paragraph{two-point function}\label{sec:eta two-point}
Let us consider the simplest case $f_1=f_2 = f$ \footnote{For the other cases $f_1\neq f_2$, we can not obtain a clear result as \eqref{eq:ACTIONwithEOW}.}. 
Using the above expressions, we can obtain
\AM{\begin{equation}
\begin{split}
	\delta^2_{j_\Delta}S&=\int \frac{d\Delta }{2\pi i}\frac{d\Delta'}{2\pi i} \int_{\eta_1-\eta_2}\hspace{-10pt} d\rho d^2\Omega ~\eta^2\sinh^2\!\rho~\eta_i^{-\Delta-\Delta'-1}\\
	&\qquad \qquad \times \int d^2z d^2 z' \phi_{\Delta'}(\rho,w,\bar{w};z',\bar{z}')\phi_\Delta(\rho,w,\bar{w};z,\bar{z})\delta j_{\Delta}^i(z,\bar{z})\frac{\delta j^i_{\Delta'}(z',\bar{z}')}{\Delta-f}\\
		&=\int \frac{d\Delta}{2\pi i}\frac{ d\Delta'}{2\pi i}\int d^2z ~\frac{\pi i}{(1-\Delta)(\Delta-f)}\delta(\Delta+\Delta'-2)\delta j^\pm_\Delta(z,\bar{z})\delta j^\mp_{\Delta'}(z,\bar{z}),
\end{split}\label{eq:ACTIONwithEOW}
	\end{equation}}
 where we used the Klein-Goldon inner product again.
 Thus we can read off the two-point function as
\begin{align}
	\langle O^+_\Delta(z,\bar{z})O^-_{\Delta'}(z',\bar{z}')\rangle &\propto \delta(\Delta+\Delta'-2)\delta^2(z-z').
\end{align}
This two-point function coincides with the celestial CFT two-point function \eqref{eq:CCFTtwopointfunction} up to the overall factors.

%
%



\section{Bulk two-point function of Massive Bulk Scalar}\label{appendix:massiveScalar}

In this appendix, we briefly comment on the derivation of a two-point function of a massive bulk scalar field by applying techniques explained in subsections \ref{subsec:CCFTvsWCFT} and \ref{subsec:changingbdyCondi}. However, we note that, for a massive bulk scalar field, we need to consider the conformal primary wavefunction for a massive field \eqref{eq:cpwForMassive}, instead of that for a massless field, \eqref{def:cpw}.

 First, in the massive case, we can expect that the action will have $O^\Delta_WO_W^\Delta$ terms, unlike the massless case \ref{subsec:changingbdyCondi}. This is the difference of $\eta$-dependence of conformal primary wavefunctions for massive and massless cases, \eqref{eq:cpwForMassive} and \eqref{def:cpw}.
 Indeed, we have\footnote{For the massless case, we have 
 \begin{equation}
	\int_{0}^{\infty} \frac{d\eta}{\eta} \eta^{\nu} \eta^{\nu'}\propto \delta(\nu+\nu').
\end{equation}
}\footnote{In this calculation, the technique from \cite{Costa:2014kfa} was used. }
\AM{\begin{align}
	\int_{0}^{\infty} \frac{d\eta}{\eta} H^{(\pm)}_{\nu}(m\eta)H_{\nu'}^{(\pm)}(m\eta)&=-\frac{2}{\pi}\Gamma[i\nu]\Gamma[-i\nu][e^{\pm\pi \nu}\delta(\nu-\nu')+\delta(\nu+\nu')],\\
 \int_{0}^{\infty} \frac{d\eta}{\eta} K_{\nu}(m\eta)K_{\nu'}(m\eta)&=8\pi\Gamma[i\nu]\Gamma[-i\nu][\delta(\nu-\nu')+\delta(\nu+\nu')],
\end{align}}
and the two delta functions $\delta(\nu + \nu')$ and $\delta(\nu-\nu')$ imply the existence of the term $O^\Delta_WO_W^\Delta$.

Since $\nu$ is restricted to $\nu\geq0$ for the massive case, it is sufficient to consider the case $\Delta=\Delta'$. Then, the action of region $\mathcal{A}^+$ is given by
\begin{equation}
\begin{split}
	S_{\mathcal{A}^+}\propto&e^{2 R}\int d^2\Omega\frac{d\Delta}{2\pi i} (e^{(\Delta-2)R}\widehat{g_{\Delta}(\partial+\gamma)g_{\Delta}}e^{(\Delta-2)R})[\Omega,\Omega_0']O_W^\nu[\Omega]O_W^\nu[\Omega'_0]\\
    &=\int d^2\Omega d^2\Omega_0'\\
    &\quad \times\left[(\gamma+i\nu-1+\epsilon)e^{(2\epsilon+2i\nu)R}\delta(\Omega-\Omega') \right.\\
    &\qquad+e^{(-2i\nu-2\epsilon)R}(\gamma-i\nu-1-\epsilon)\frac{i\nu}{4\pi}\frac{i\nu}{4\pi}\int d\Omega_0 \left(\frac{2}{1-\cos\tau}\right)^{1+i\nu}\!\left(\frac{2}{1-\cos\tau'}\right)^{1+i\nu}\\
	&\qquad +(\gamma-i\nu-1-\epsilon)\frac{i\nu}{4\pi}\left(\frac{2}{1-\cos\tau}\right)^{1+i\nu}\\
 &\hspace{4cm}+\left.\!(\gamma+i\nu-1+\epsilon)\frac{i\nu}{4\pi}\left(\frac{2}{1-\cos\tau}\right)^{1+i\nu}\right]O_W^\nu[\Omega]O_W^\nu[\Omega'_0],
\end{split}\label{eq:MASSIVEactionOnAdS}
	\end{equation}
where  the factors appearing in $(e^{(\Delta-2)R}\widehat{g_{\Delta}(\partial+\gamma)g_{\Delta}}e^{(\Delta-2)R})[\Omega,\Omega_0']$ is the same as those appearing in \ref{subsec:changingbdyCondi}.

We can compare this action \eqref{eq:MASSIVEactionOnAdS} with the massless case \eqref{eq:MasslessActionOnAdSPatch}.
For the massless case, we had to renormalize by  setting $\gamma=1$ to remove the delta function in \eqref{eq:MasslessActionOnAdSPatch}. On the other hand, for massive case \eqref{eq:MASSIVEactionOnAdS}, we do not have to  do such an operation\footnote{Here, we note that at $\nu=0$, we can not use the rapid oscillating argument. 
Here we just assume that we can ignore such contribution here since we will need special treatment on such a point. In the near future, we will revisit such contributions.
}. This is because, in the action \eqref{eq:MASSIVEactionOnAdS}, the phase factor of the first term rapidly oscillates at large $R$ and the first term vanishes. The resulting two-point function coincides with the one obtained from Celestial CFT \eqref{eq:massiveCCFTtwopoint} up to the overall factor. This is consistent with the result obtained in \cite{Ogawa:2022fhy}.


\bibliographystyle{JHEP}
\bibliography{reference}

\end{document}